\documentclass[twocolumn,preprintnumbers]{revtex4}
\usepackage{graphicx}
\usepackage{amsmath}
\usepackage{amssymb}
\usepackage{amsthm}
\usepackage{amsfonts}
\usepackage{enumerate}
\usepackage{centernot}
\usepackage{tikz}

\begin{document}

\title{Massive Spin-$3/2$ Rarita-Schwinger Quasiparticle in Condensed Matter
Systems}
\author{Feng Tang$^{1,4}$}
\thanks{These two authors contribute equally.}
\author{Xi Luo$^{2}$}
\thanks{These two authors contribute equally.}
\author{Yongping Du$^{5,1,4}$}
\author{Yue Yu$^{3,4}$}
\author{Xiangang Wan$^{1,4}$}
\affiliation{${}^{1}$National Laboratory of Solid State Microstructures and School of
Physics, Nanjing University, Nanjing 210093, China}
\affiliation{${}^{2}$CAS Key Laboratory of Theoretical Physics, Institute of Theoretical
Physics, Chinese Academy of Sciences, Beijing 100190, China}
\affiliation{${}^{3}$State Key Laboratory of Surface Physics, Center for Field Theory and
Particle Physics, Department of Physics, Fudan University, Shanghai 200433,
China}
\affiliation{${}^{4}$Collaborative Innovation Center of Advanced Microstructures, Nanjing
210093, China}
\affiliation{${}^{5}$Department of Applied Physics, Nanjing University of Science and
Technology, Nanjing 210094, China}

\begin{abstract}
The spin-$3/2$
elementary particle, known as Rarita-Schwinger (RS) fermion, is described by a vector-spinor field $\psi _{\mu \alpha }$,
whose number of components is larger than its independent degrees of freedom
(DOF). Thus the RS equations contain nontrivial constraints to eliminate the
redundant DOF. Consequently the standard procedure adopted in realizing
relativistic spin-1/2 quasi-particle is not capable of creating the RS
fermion in condensed matter systems. In this work, we propose a generic
method to construct a Hamiltonian which implicitly contains the RS
constraints, thus includes the eigenstates and energy dispersions being
exactly the same as those of RS equations. By implementing our $16\times 16$
or $6\times 6$ Hamiltonian, one can realize the 3 dimensional or 2
dimensional (2D) massive RS quasiparticles, respectively. In the
non-relativistic limit, the 2D $6\times 6$ Hamiltonian can be reduced to two
$3\times 3$ Hamiltonians which describe the positive and negative energy
parts respectively.
Due to the nontrivial constraints, this simplified 2D massive RS
quasiparticle has an exotic property: it has vanishing orbital magnetic
moment while its orbital magnetization is finite. Finally, we discuss the
material realization of RS quasiparticle. Our study provides an opportunity to realize higher spin elementary fermions with constraints in condensed matter systems. 
\end{abstract}

\date{\today}
\maketitle

\section{Introduction.}

By requiring the Lorentz invariance and a positive probability density,
Dirac proposed in 1928 a relativistic wave equation \cite{Dirac}. Describing
spin-$1/2$ particles, this equation is known as the Dirac equation \cite%
{Dirac}. In 1929, Weyl introduced Weyl equation, a simplified version of the
Dirac equation for relativistic particles, whose solutions predicted
massless fermions with a definite handedness (or chirality) \cite{Weyl-1929}%
. The third relativistic quantum {mechanical} equation for spin-$1/2$
fermion is Majorana equation, and the antiparticle of Majorana fermion is
itself \cite{Majorana}. It is well known that Dirac equation describes
electrons etc. However, Weyl and Majorana fermions have not been observed as
{an elementary} particle in the laboratory or nature {yet}. In spite of the
huge energy-scale difference, currently the rich correspondences between
high energy physics and condensed matter (CM) physics attract a lot of
research attentions. Several unique electronic structures in certain CM
systems (CMS) behave analogously to elementary particles. For instance, the
Dirac quasi-fermions in graphene \cite{Dirac,graphene,Graphene-2,Graphene-3}
and Dirac semimetals \cite{DSM-1,DSM-2,DSM-3}. The Weyl quasiparticle was
proposed \cite{WSM-1,Balents,WSM-2,WSM-3,WSM-4,WSM-4-2} and has been
experimentally confirmed in Weyl semimetals \cite{WSM-5,WSM-6,WSM-7}. The
Majorana zero modes in CMS were also discussed intensively \cite{MR,RG,Ki}.

It is worth emphasizing that the properties of these quasiparticles in CMS
are much more flexible and fruitful than their high energy counterparts: (i)
The finite size of the system gives these quasiparticles the chance to see
the peculiar edge and surface states of the world \cite{edge,WuBZ,WSM-1}.
(ii) The minimal spacing of the lattice makes them safe from the ultraviolet
divergence. (iii) Novel tunneling properties associated with Klein paradox
may be experimentally examined \cite{Graphene-3}. (iv) The effect of the
chiral anomaly can be evaluated by highly anisotropic negative
magnetoresistance \cite{Chiral-1,Chiral-2}. (v) There are quasiparticles
with non-abelian statistics \cite{MR}, which may be applied to topological
quantum computation \cite{Ki1}.

Without exhaustively listing the intriguing features of the spin-$1/2$
relativistic quasiparticles in CMS, however, the basic properties of these
spin-$1/2$ fermions have been adequately studied and well-understood. On the
other hand, the {relativistic} particles with spin higher than one appear
neither as elementary particles nor as quasiparticles in CMS, although they
were studied in the early day of quantum mechanics \cite{Dirac1}.
Especially, the spin-$3/2$ fermion, which obeys Rarita-Schwinger (RS)
equations \cite{RS-Equation}, plays a key role in supergravity theory \cite%
{F-N}. Very recently, by considering the fourfold degenerate band crossing
points, the possible spin-$3/2$ generalization of the Dirac equation in CMS
were proposed \cite{LY,IF,newF,Ez}. Instead of the shackle from Lorentz
invariance in quantum field theory, these quasiparticles are only subject to
the crystallographic symmetry groups. Therefore, one can classify the band
crossing points which are stabilized by space group symmetries and then
search for the possible quasi-fermions {that have no} counterparts in high
energy physics \cite{newF,Nature-Ding-H,New-1,New-2,New-3,New-4,New-5}.

Among the interesting properties of the RS fermion, the most astonishing one
is that external magnetic field can induce a superluminal mode for the
massive RS (M-RS) fermion \cite{mass2,mass3}. The \textquotedblleft speed of
light\textquotedblright\ which is usually the Fermi velocity in CMS is not
the upper limit of the quasiparticle velocity and thus the \textquotedblleft
superluminal\textquotedblright\ modes and \textquotedblleft
acausality\textquotedblright\ are basically permitted. The two dimensional
(2D) M-RS gas also shows a nonlinear quantum Hall phenomenon \cite{lfwy}
which is different from the well-known quantum Hall effects in the 2D
electron gas \cite{qhe2deg} and the Dirac fermion gas in graphene \cite%
{QHE-Graphene}. Therefore, to create the M-RS quasiparticles in CM
playground is a very intriguing topic, which we address in this work. {%
Inspired by the CM simulation of the Dirac fermion\cite{semenoff}, we aim to
construct a Hamiltonian with }the wave function and the corresponding energy
dispersion around the Fermi {level approximately the same} as the
relativistic spin-$3/2$ RS particles. Thus the low-lying excitations are the
RS quasiparticles.

In high energy physics, it is convenient to use either equation of motion or
Lagrangian, while Hamiltonian {formalism is favored} in CMS. {To obtain the
corresponding Hamiltonian for relativistic spin-$1/2$ particles is
straightforward because their wave equations are the ones with first order
time derivative.} By requiring that the effective electronic Hamiltonian
around some points in momentum space of CMS approximately the same as the
Hamiltonian in high energy physics, one can realize spin-$1/2$ relativistic
fermions (such as Dirac, Weyl and Majorana ferimon) in CMS as emergent
phenomena. However, the wave function of spin-$3/2$ M-RS fermions,
represented by a vector-spinor $\psi _{\mu }$, are of $16 $ or $6$
components for $d$ ($d=3$ or $2$) dimensional systems, while the number of
the genuine degrees of freedom (DOF) are only $8$ or $2$ \cite{SM}. Thus, in
addition to first order time derivative equations, RS equations also contain
constraint equations, which do not exist for spin-1/2 fermions, to eliminate
the redundant DOF.

The Hamiltonian for high-spin relativistic fermions has also been derived
\cite{Hamil}. However in addition to the Hamiltonian \cite{Hamil}, which is
only for the genuine DOF, one still needs to explicitly include the
constraint equations to obtain all the components of RS wave function $\psi
_{\mu }$ to completely solve the RS particle problem as shown in the
following section. Thus one cannot simply use this Hamiltonian\cite{Hamil}
and follows the standard procedure adopted in spin-$1/2$ ferimons to realize
the spin-$3/2$ quasiparticle in CMS. In order to realize relativistic RS
excitations in CMS, in this work for 3D (or 2D) we construct a $16$-band (or
$6$-band) Hamiltonian which implicitly includes the RS constraint
conditions. Such a Hamiltonian contains the eigen-solutions which are
exactly the same as those of RS fermion. Since the dimension of our
Hamiltonian is larger than the genuine DOF of the RS fermion, thus in
addition to the RS\ modes, the proposed Hamiltonian also contains non-RS
modes. We prove that these RS modes can be separated from the non-RS modes.
By implementing our proposed 16$\times $16 or 6$\times $6 Hamiltonians in
CMS, the 3D or 2D massive relativistic spin-$3/2$ RS quasiparticles can be
realized. Focusing on the 2D case, which is much easier to realize in
realistic materials, we find that in the {non-relativistic} simplification
with wavefunction approximated up to the first order of $p/m$, the 6$\times $%
6 Hamiltonian can be reduced to two $3\times 3$ matrices. We also find a
novel property of these non-relativistic RS modes compared with those
described by the Schr\"{o}dinger and Dirac equations. In general, the Berry
curvature and orbital magnetic moment formally have the same properties
under the symmetry group. However for the 2D {non-relativistic} M-RS
quasiparticles, we find that due to the nontrivial constraints on the RS
quasiparticle, the Berry curvature is finite while the orbital magnetic
moment vanishes. We also derive a sufficient condition which will restrict
the orbital moment to be exactly zero. At last we search for the {2D
non-relativistic }M-RS quasiparticles around particular points in the
Brillouin zone (BZ) and suggest the layered structures constructed from
several trigonal and hexagonal lattices can hold such a low energy
excitation. As an example, based on \emph{ab initio} calculations, we
predict that CaLiX (X=Ge and Si) are the candidates.

The paper is organized as follows. In Sec. II, we will present the
Hamiltonian formalism of the massive high energy RS fermion in 2D and 3D.
Realizing the non-trivial constraints of the RS fermions is the most
difficult obstacle in CMS, and we propose a generic method to implicitly
include these constraints in the effective Hamiltonian in Sec. III. {We also
give the 16}$\times ${16 and 6}$\times ${6 Hamiltonians which contain modes
with the wavefunctions and energy spectra being exactly the same as the RS
equations\ in Sec. III, so that one can realize 3D and 2D M-RS
quasiparticles by implementing these Hamiltonian in CMS.} {Focusing on 2D
case, we prove that in the non-relativistic approximation with wavefunction
up to the first order of }$p/m${, the positive and negative energy parts of
RS wavefunctions decouples completely. Consequently the effective
Hamiltonian for 2D RS quasiparticles can be reduced two 3}$\times ${3
Hamiltonians in Sec. IV. In Sec. V, }we show that due to the nontrivial
constraints, this simplified 2D massive RS quasiparticle has an exotic
property: although the intrinsic orbital magnetic moment of an energy band
transforms like its Berry curvature under symmetry operations, the former is
exactly zero-valued while the latter is finite. We also propose a possible
material for the {non-relativistic RS} insulator in Sec. VI. Finally we
summarize our results, comments on the parton construction and the
topological classification of the non-relativistic RS insulator in the
section of Discussions and Conclusions.

\section{Hamiltonians of RS fermion for $d=3$ and $2$ in high energy physics.%
}

{The Lagrangian describing the massive relativistic spin-$\frac{3}{2}$
particles was first proposed by Rarita and Schwinger \cite{RS-Equation}. The
corresponding equation of motion reads
\begin{eqnarray}
&&(\gamma \cdot \partial +m)\psi _{\mu\alpha }=0,  \label{RS-Equ} \\
&&\gamma \cdot \psi =\gamma_\mu\psi_\mu=0,  \label{RS-Constr}
\end{eqnarray}%
which is a special case of Bargmann-Wigner (B-W) equation \cite{BW-Equ} that
works for any arbitrary spin.} The vector-spinor $\psi _{\mu\alpha}$ is of $%
16$ or $6$-components for $d=3$ or $2$, {$\mu$ is the vector index while $%
\alpha$ is the spinor index, and we choose the Dirac representation for the $%
\gamma$ matrices and will omit the spinor index for later convenience. }For
the detailed description of the notations, see Secs. I and II of the SM \cite%
{SM}. 
With Eqs. (\ref{RS-Equ}) and (\ref{RS-Constr}), one can obtain a subsidiary
condition $\partial \cdot \psi =0$. This constraint and Eq. (\ref{RS-Constr}%
) project out the redundant spin-$1/2$ DOF and leave the genuine spin-$3/2$
ones.

{In CMS, the momentum-space language is preferred}. Thus we rewrite the RS
equations (\ref{RS-Equ}) and (\ref{RS-Constr}) in the momentum space%
\begin{align}
i\partial _{t}\psi _{\mu }(\mathbf{p},t)& =(\boldsymbol{\alpha }\cdot
\mathbf{p}+\beta m)\psi _{\mu }(\mathbf{p},t),  \label{RS-Motion} \\
\psi _{d+1}(\mathbf{p},t)& =i\alpha _{i}\psi _{i}(\mathbf{p},t),
\label{RS-p}
\end{align}%
where $\boldsymbol{\alpha }=i\gamma _{d+1}\boldsymbol{\gamma }$ and $\beta
=\gamma _{d+1}$. The constraint (\ref{RS-p}) implies that the RS fields are
not just \textit{d}+1 copies of Dirac fields. Under SO($d$) rotations, the
time-like spinor $\psi _{d+1}$ transforms as the $j=\frac{1}{2}$ irreducible
representation, while the space-like spinors ${\psi _{i}}^{\prime }s\ $(i.e.
${\psi }_{x},{\psi }_{y},{\psi }_{z}$ for $d=3$ or $\psi _{x},\psi _{y}$ for
$d=2$) transform as a product representation: $(j=1)\otimes (j=\frac{1}{2})$
reducible representation which contains the irreducible ones $\phi _{3/2}(j=%
\frac{3}{2})$\ and $\phi _{1/2}(j=\frac{1}{2})$\ \cite{note}. The details of
the transformation from ${\psi _{i}}^{\prime }s$\ to $\phi _{3/2}$\ and $%
\phi _{1/2}$ are shown in Secs. III and IV of the SM \cite{SM}. Eq. (\ref%
{RS-Motion}) is then rewritten as
\begin{align}
& i\partial _{t}\phi _{3/2}(\mathbf{p},t)=h_{33}(\mathbf{p})\phi _{3/2}(%
\mathbf{p},t)+h_{31}\phi _{1/2}(\mathbf{p},t),  \label{RS3-1} \\
& i\partial _{t}\phi _{1/2}(\mathbf{p},t)=h_{13}(\mathbf{p})\phi _{3/2}(%
\mathbf{p},t)+h_{11}\phi _{1/2}(\mathbf{p},t),  \label{RS1-1} \\
& i\partial _{t}\psi _{d+1}(\mathbf{p},t)=H_{t}(\mathbf{p})\psi _{d+1}(%
\mathbf{p},t),  \label{RST}
\end{align}%
where $H_{t}=\boldsymbol{\alpha }\cdot \mathbf{p}+\beta m$.
The constraint equation Eq. (\ref{RS-p}) can be rewritten as the following:%
\begin{equation}
\psi _{d+1}(\mathbf{p},t)=R_{t1}\phi _{1/2}(\mathbf{p},t),
\label{new-constrain}
\end{equation}%
with $R_{t1}$ being a constant matrix, e.g., $R_{t1}=-i\sqrt{2}$ for \textit{%
d}=2 ($R_{t1}$ for \textit{d}=3 is given in Sec. {III} of the SM \cite{SM}).
{Substituting Eq. (\ref{new-constrain}) into Eq. (\ref{RST}), we obtain the
Hamiltonian $H_{1/2}$ for $\phi_{1/2}$,}
\begin{eqnarray}
i\partial _{t}\phi _{1/2} &=&H_{1/2}\phi _{1/2},  \label{1/2-Equ} \\
H_{1/2} &=&R_{t1}^{-1}H_{t}R_{t1}.
\end{eqnarray}%
With {Eq. (\ref{1/2-Equ})} and (\ref{RS1-1}), the relation between $\phi
_{1/2}$ and $\phi _{3/2}$ is given by
\begin{equation}
\phi _{1/2}(\mathbf{p},t)=R_{13}(\mathbf{p})\phi _{3/2},  \label{Con-1}
\end{equation}%
where, e.g., $R_{13}=\frac{1}{p^{2}+4m^{2}}[2mp_{1}+2mip_{2}\sigma
_{3}-i(p_{1}^{2}-p_{2}^{2})\sigma _{2}+2ip_{1}p_{2}\sigma _{1}]$ for $d=2$. (%
$R_{13}$ for $d$=3 is given in Sec. III of the SM \cite{SM}.) {Substituting
this relation (\ref{Con-1}) into Eq. (\ref{RS3-1}), we reveal the genuine
Hamiltonian $H_{3/2}=h_{33}+h_{31}R_{13}$ for $\phi_{3/2}$, }
\begin{widetext}
\begin{eqnarray}
i\partial _{t}\phi _{3/2}(\mathbf{p},t) &=&H_{3/2}\phi _{3/2}(\mathbf{p},t),
\label{H-3/2} \\
H_{3/2} &=&(m+\frac{2mp^{2}}{4m^{2}+p^{2}})\sigma _{3}+\frac{%
p_{1}(3p_{2}^{2}-p_{1}^{2})}{4m^{2}+p^{2}}\sigma _{1}+\frac{%
p_{2}(p_{2}^{2}-3p_{1}^{2})}{4m^{2}+p^{2}}\sigma _{2},\text{ for }d=2; \\
H_{3/2} &=&\sigma _{1}\otimes (\mathbf{\Sigma }^{\frac{3}{2}}\cdot \mathbf{p}%
)[1+\frac{p^{2}-(\mathbf{\Sigma }^{\frac{3}{2}}\cdot \mathbf{p})^{2}}{\frac{4%
}{9}p^{2}+m^{2}}]+m\sigma _{3}\otimes \lbrack 1+\frac{1}{2}\frac{p^{2}-(%
\mathbf{\Sigma }^{\frac{3}{2}}\cdot \mathbf{p})^{2}}{\frac{4}{9}p^{2}+m^{2}}],%
\text{ for }d=3,  \label{3D-H-3/2}
\end{eqnarray}%
\end{widetext}
where
$\mathbf{\Sigma}_{\frac{3}{2}}= \frac{2}{3}\mathbf{J}_{\frac{3}{2}}$, and $%
\mathbf{J}_{\frac{3}{2}}$ is angular momentum matrix in the $j=3/2$
representation of SO(3). Eq. (\ref{3D-H-3/2}) is consistent with that
derived by Moldauer and Case \cite{Hamil} as expected.

\section{Construction of Hamiltonian for realizing RS quasiparticle in CMS.}

Suppose $\phi _{3/2}$ is the eigenvector of $H_{3/2}$ {with the eigen-energy
$\epsilon $, then the components $\phi _{1/2}$\ and $\psi _{d+1}$ in the RS
wave function ($\phi _{3/2},\phi _{1/2},\psi _{d+1}$)$^{T}$ can be obtained
through the constraints Eq.(\ref{Con-1}) and Eq.(\ref{new-constrain}). They
share the same eigen-energy $\epsilon $, namely,}
\begin{eqnarray}
H_{3/2}\phi _{3/2} &=&\epsilon \phi _{3/2},  \label{3/2} \\
H_{1/2}\phi _{1/2} &=&\epsilon \phi _{1/2},  \label{1/2} \\
H_{t}\psi _{d+1} &=&\epsilon \psi _{d+1}.  \label{d+1}
\end{eqnarray}%
{Therefore } combining with the constraints Eq. (\ref{Con-1}) and Eq. (\ref%
{new-constrain}), the Hamiltonian $H_{3/2}$ is sufficient {for obtaining the
RS wave function}, and eventually the equation of motion of the RS fermion
can be solved. However, this field theory approach is not operable for CM
simulation {because in addition to the Hamiltonian one cannot explicitly add
any other equations.}

In order to obtain the RS-like excitations in CMS, we propose a generic
method of constructing a $16$ or $6$-band Hamiltonian $\bar{H}(\mathbf{p})$,
which contains eigenvector $\Phi \ $with components $(\Phi _{3/2},\Phi
_{1/2},\Phi _{d+1})^{T}$ being exactly the same as the vector-spinor {eigen
wave function} ($\phi _{3/2},\phi _{1/2},\psi _{d+1}$)$^{T}$ of RS
equations, with the same eigenvalue $\epsilon $, i.e. $\bar{H}\Phi =\epsilon
\Phi $, written in the following block form,
\begin{equation}
\left(
\begin{array}{ccc}
\bar{H}_{33} & \bar{H}_{31} & \bar{H}_{3t} \\
\bar{H}_{31}^{\dag } & \bar{H}_{11} & \bar{H}_{1t} \\
\bar{H}_{3t}^{\dag } & \bar{H}_{1t}^{\dag } & \bar{H}_{tt}%
\end{array}%
\right) \left(
\begin{array}{c}
\Phi _{3/2} \\
\Phi _{1/2} \\
\Phi _{d+1}%
\end{array}%
\right) =\epsilon \left(
\begin{array}{c}
\Phi _{3/2} \\
\Phi _{1/2} \\
\Phi _{d+1}%
\end{array}%
\right) .  \label{H-CM}
\end{equation}

As $\Phi $ is the same as the eigenvector of the RS equations, $\Phi _{3/2}$%
,\ $\Phi _{1/2}$ and $\Phi _{d+1}$ should satisfy the Eq. (\ref{3/2}), Eq. (%
\ref{1/2}) and Eq. (\ref{d+1}), respectively, meanwhile they must also obey
the RS constrains of Eq. (\ref{Con-1}) and Eq. (\ref{new-constrain}). Then
we can take $\bar{H}_{33}=H_{3/2},$\ $\bar{H}_{11}=H_{1/2}$ and $\bar{H}%
_{tt}=H_{t}$ for the diagonal terms and {demand} the off-diagonal parts
to satisfy the following three relations:
\begin{eqnarray}
&&\bar{H}_{31}R_{13}+\bar{H}_{3t}R_{t1}R_{13}=0,  \label{H-CM-condition1} \\
&&\bar{H}_{31}^{\dag}+\bar{H}_{1t}R_{t1}R_{13}=0,  \label{H-CM-condition2} \\
&&\bar{H}_{3t}^{\dag}+\bar{H}_{1t}^{\dag}R_{13}=0.  \label{H-CM-condition3}
\end{eqnarray}%

According to Eqs. (\ref{H-CM-condition2}) and (\ref{H-CM-condition3}), $\bar{H}_{31}$ and $\bar{H}_{3t}$ can be expressed by $\bar{H}_{1t}$. Furthermore {\ if $\bar{H}_{1t}$} satisfies
\begin{equation}
R_{t1}^{\dag }\bar{H}_{1t}^{\dag }=-\bar{H}_{1t}R_{t1},  \label{R_t1}
\end{equation}%
the requirement namely Eq. (\ref{H-CM-condition1}) is automatically fulfilled. In high energy
physics, {the number of linearly independent solutions} of the RS equations
is exactly equal to the number of the genuine DOF, namely all of those {%
linearly independent solutions form the basis for quantizing the RS field.}
However in CMS, the Hamiltonian $\bar{H}$ has {16 (6) linearly independent
eigen-solutions in 3D (2D)}, of which 8 (2) are {identified as the emergent
RS modes while the non-RS solutions can be separated from the RS ones which
is one of the merits of our method. Because of the free parameters in the
off-diagonal terms of $\bar{H}$, we are able to tune the spectrum of the
non-RS modes away from the RS ones. We will explain more details on this in
the following.}

\begin{figure}[tbp]
\centering
\includegraphics[width=6cm]{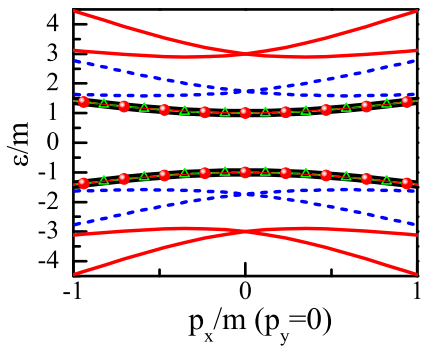}
\caption{For \textit{d}=2, We plot the six bands from the $\bar{H}$\ in Eq. (%
\protect\ref{H-CM}) (namely 2D M-RS Hamiltonian in CMS). For $H_{1t}=m(%
\protect\sigma _{1}+\protect\sigma _{2})$, the green triangles and dashed
blue lines represent the RS and non-RS modes, repectively. We also calculate
the selection of $H_{1t}=2m(\protect\sigma _{1}+\protect\sigma _{2})$, and
plot the RS and non-RS modes by the red solid dots and lines, respectively.\
For comparison, we also show the dispersions of RS fermion by the two black
thick lines. It is clear that the RS modes are exactly the same as the RS
fermion, meanwhile the RS modes are separated from the non-RS modes. }
\label{rsbands}
\end{figure}

{Taking \textit{d}=2 CMS for example, the general form of the off-diagonal
term $H_{1t}$ satisfying Eq. (\ref{R_t1}) reads
\begin{equation}
\bar{H}_{1t}=\left(
\begin{array}{cc}
a_{1} & b_{-} \\
b_{+} & a_{2}%
\end{array}%
\right) ,
\end{equation}%
with $b_{\pm }=b_{1}\pm ib_{2}$ and $a_{1},a_{2},b_{1},b_{2}\in \mathbb{R}$
being free parameters. Once $H_{1t}$ is given, the other off-diagonal terms $%
\bar{H}_{13}$ and $\bar{H}_{t3}$ are generated through Eq. (\ref%
{H-CM-condition2}) and (\ref{H-CM-condition3}), thus the Hamiltonian $\bar{H}
$ is determined. In other words, $\bar{H}$ is parametrized by $%
a_{1},a_{2},b_{1},$ and $b_{2}$. We plot the spectrum of $\bar{H}$ in Fig. %
\ref{rsbands}. Two of the bands are the RS bands while the other four are
the non-RS ones. They can be separated by tuning the free parameters as
shown in Fig. \ref{rsbands}. Meanwhile for any arbitrary choice of the
parameters, the RS modes are unchanged with the eigen-energies and the wave
functions being identical to those of the RS fermions in high energy
physics. } %
%
%
%
%

\section{Non-relativistic approximation for Hamiltonian of CMS.}

With the same wave functions of RS fermion in the low-lying excitation
sector, however, the Hamiltonian $\bar{H}$ 
is not directly applicable to real CM materials because of its nonlocal
elements arising from, say, the factor $\frac{1}{p^{2}+4m^{2}}$ in $\bar{H}%
_{33}$ for $d=2$. Due to a finite mass, we make the series expansion of the
nonlocal terms in $\bar{H}$ in terms of $\frac{\mathbf{p}}{m}$. The
Hamiltonian is then expanded as a $16\times 16$ matrix for $d=3$ shown in
Sec. VII of the SM \cite{SM}, and $6\times 6$ matrix
for $d=2$ as follows,
\begin{widetext}
\begin{equation}\label{h6by6}
\bar{H}\sim
\left(\begin{array}{cccccc}
m+\frac{p^2}{2m}& 0& -i\frac{a_1p_-}{\sqrt2m}& -i\frac{b_-p_-}{\sqrt2m}& -\frac{a_1p_-}{2m}& -\frac{b_-p_-}{2m}\\
0& -m-\frac{p^2}{2m}&  -i\frac{b_+p_+}{\sqrt2m}& -i\frac{a_2p_+}{\sqrt2m}& -\frac{b_+p_+}{2m}& -\frac{a_2p_+}{2m}\\
i\frac{a_1p_+}{\sqrt2m}& i\frac{b_-p_-}{\sqrt2m}& m& p_-& a_1& b_-\\
i\frac{b_+p_+}{\sqrt2m}& i\frac{a_2p_-}{\sqrt2m}& p_+& -m& b_+& a_2\\
-\frac{a_1p_+}{2m}& -\frac{b_-p_-}{2m}& a_1& b_-& m& p_-\\
-\frac{b_+p_+}{2m}& -\frac{a_2p_-}{2m}& b_+& a_2& p_+& -m
\end{array}\right).
\end{equation}
\end{widetext}By implementing this Hamiltonian, one can realize the 2D
massive relativistic spin-$3/2$ RS quasiparticles in CMS. Furthermore we
notice that in the non-relativistic limit, the Hamiltonian (\ref{h6by6}) can
be further simplified which makes it much easier to realize in realistic CM
materials.

{\ Up to the first order of $\frac{\mathbf{p}}{m}$, the wave functions of
the RS bands read
\begin{eqnarray}
&&\Psi_+=(1,0,\frac{p_{x}+ip_{y}}{2m},0,-i\frac{p_{x}+ip_{y}}{ \sqrt{2}m}%
,0)^{T}, \\
&&\Psi_-= (0,1,0,\frac{p_{x}-ip_{y}}{2m},0,-i\frac{ p_{x}-ip_{y}}{\sqrt{2}m}%
)^{T},
\end{eqnarray}
with the corresponding eigen energies $\epsilon_\pm=\pm(m+\frac{p^2}{2m})$.
These spectra and wave functions are exactly identical to those of the RS
fermions in the non-relativistic limit. Furthermore, }
%
the positive and negative energy sectors are completely separated. Therefore
the Hamiltonian in CMS can be reduced to a block diagonal form {up to a
unitary transformation. One is for particle while the other is for hole
which are denoted as $\tilde{H}_{+}$ and $\tilde{H}_{-}$ respectively}. The
Hamiltonians then read as\cite{SM}
\begin{eqnarray}
&&\tilde{H}_{\pm }(\mathbf{q})=  \label{Hk.p} \\
&&\left(
\begin{array}{ccc}
\Delta _{1}(q_{x}^{2}+q_{y}^{2}) & \Delta _{2}q_{\mp } & \Delta _{2}q_{\mp }
\\
\Delta _{2}q_{\pm } & \Delta _{1}(q_{x}^{2}+q_{y}^{2})+\Delta _{3} & 0 \\
\Delta _{2}q_{\pm } & 0 & \Delta _{1}(q_{x}^{2}+q_{y}^{2})-\Delta _{3}%
\end{array}%
\right) ,  \notag \\
&&  \notag
\end{eqnarray}%
where $\mathbf{p}$ is replaced by $\mathbf{q}=(q_{x},q_{y})$ which is the
displacement away from a particular point in the BZ (RS point), $q_{\pm
}=q_{x}\pm iq_{y}$. The real parameters $\Delta _{1}$,\ $\Delta _{2}$\ and $%
\Delta _{3}$\ can be used to construct a \textquotedblleft light
velocity\textquotedblright\ $c^{\prime }$: $c^{\prime }=|\sqrt{\frac{3}{2}}%
\frac{\Delta _{1}\Delta _{3}}{\Delta _{2}\hbar }|$.

To summarize this section, we shall remind that the Hamiltonian (\ref{Hk.p})
effectively describes the non-relativistic 2D M-RS quasiparticles with the
eigen wave functions being identical to the 2D M-RS solution to the first
order of $p/m$. As a self-consistency check, this non-relativistic 2D M-RS
quasiparticle also satisfies the RS\ constraints Eq. (\ref{Con-1}) and Eq. (%
\ref{new-constrain}) to the first order of $p/m$. As addressed in the next
section, the non-relativistic M-RS wave function possesses a novel propertie
which have not been found for spin-1/2 non-relativistic particles.


\section{Exotic propety: finite orbital magnetization with vanishing orbital
magnetic moment.}

The non-relativistic 2D M-RS quasiparticles possesses exotic Berry-phase
related properties {which can be seen by solving $\tilde{H}_{+}(\mathbf{q})$%
\ in Eq. (\ref{Hk.p}) directly (the case of $\tilde{H}_{-}(\mathbf{q})$ is
similar).} Three eigenvalues of $\tilde{H}_{+}(\mathbf{q})$ are $\epsilon
_{RS}=\Delta _{1}q^{2}$ (RS band) and $\epsilon _{\pm }=\Delta _{1}q^{2}\pm
\sqrt{\Delta _{3}^{2}+2\Delta _{2}^{2}q^{2}}$ (non-RS bands). The
corresponding eigenvectors are then, $u_{RS}=N_{0}(1,-\frac{\Delta _{2}}{%
\Delta _{3}}q_{+},\frac{\Delta _{2}}{\Delta _{3}}q_{+})^{T}$, $u_{+}=N_{+}(%
\frac{2\Delta _{2}q_{-}}{\Delta _{3}+\lambda (q)},1,\frac{\lambda (q)-\Delta
_{3}}{\lambda (q)+\Delta _{3}})^{T}$ and $u_{-}=N_{-}(\frac{2\Delta _{2}q_{-}%
}{\Delta _{3}+\lambda (q)},-\frac{\lambda (q)-\Delta _{3}}{\lambda
(q)+\Delta _{3}},-1)^{T}$ where $\lambda (q)=\sqrt{\Delta _{3}^{2}+2\Delta
_{2}^{2}q^{2}}$ and $N_{0},N_{\pm } $ are the normalization factors. {%
Calculating the Berry curvature is straightforward once we have the
non-degenerate eigenvalues and eigenstates. Berry curvature} %
%
behaves like a pesudo-magnetic field in the momentum space \cite{rmp}. {%
Similar to Dirac fermion}, the Berry curvature of RS band, $\Omega
_{RS}^{z}(q)=-\frac{4\Delta _{2}^{2}\Delta _{3}^{2}}{(2\Delta
_{2}^{2}q^{2}+\Delta _{3}^{2})^{2}}$, is also peaked around $q$=0 point.
However, with increasing value of $q$, it decays much faster than that of
Dirac fermion \cite{rmp}.

{Another Berry curvature related quantity is the orbital magnetic moment.
They behave exactly the same under symmetry groups\cite{rmp}.}
Arising from the self-rotating motion of the wavepacket {which describes the
dynamics of carriers in CMS}, the orbital magnetic moment plays an important
role in anomalous electronic and thermal transport properties \cite%
{rmp,W-Yao}. The orbital magnetic moment for a given band $n$ is {determined}
by $m_{n}^{z}(\mathbf{q})=-\frac{ie}{2\hbar }(<\partial _{q_{x}}u_{n}|H(%
\mathbf{q})-\epsilon _{n}(\mathbf{q})|\partial _{q_{y}}u_{n}>-c.c.)$ \cite%
{PRL-2005-1,PRL-2005-2}. We find that while the non-RS bands have finite
orbital magnetic moments ($m_{+}^{z}(q)=-m_{-}^{z}(q)=\frac{e\Delta
_{2}^{2}\Delta _{3}^{2}}{(\Delta _{3}^{2}+2\Delta _{2}^{2}q^{2})^{3/2}}$),
the orbital magnetic moment of the RS quasiparticle is exactly zero for an
arbitrary $q$ (i.e. $m_{RS}^{z}(q)=0$). This peculiar zero result is
guaranteed by the constraints on the RS fermion, not by any symmetry, {%
because the Berry curvature is finite. As mentioned before, they transform
in the same manner under symmetry groups. If this zero result of the orbital
magnetic moment is dictated by symmetry, then the Berry curvature should
also be zero.} %
%
Although the orbital magnetic moment is exactly zero, the orbital
magnetization $M$\ of the RS band is still finite due to the topological
contribution coming from the Berry phase correction to the phase-space
density of states for Bloch state\cite{rmp,PRL-2005-1,PRL-2005-2}. Thus the {%
non-relativistic} RS band possesses a remarkable feature: Despite the
nonzero orbital magnetization, under an external magnetic field there is
only spin-related energy shift or splitting in the RS band spectrum. {The
non-relativistic} RS band also provides a good platform to unambiguously
clarify the unique effect from the topological orbital magnetization\cite%
{rmp}, which usually mixes with the orbital magnetic moment thus is hard to
be identified. Having the non-zero Berry curvature, finite orbital
magnetization and vanishing orbital magnetic moment, the {non-relativistic}
RS quasiparticle thus may display not only fundamental interesting {physics}
but also peculiar properties which cannot be realized in conventional
quasiparticles.

As an important physical quantity, orbital magnetic moment describes the
intrinsic property of the carriers in a Bloch band. We give a criterion for
the vanishing orbital magnetic moment: One can always rewrite an arbitrary $%
s\times s$ Hamiltonian $H_{s\times s}(q)$ ($s=2j+1$, $j$ is an integer) as $%
H_{s\times s}(q)=h_{s\times s}(q)+C(q)I_{s\times s}$ where $C(q)$ is a
scalar quantity, while $I_{s\times s}$\ is a unit matrix. If the eigenvalues
of $H_{s\times s}(q)$\ are nondegenerate meanwhile one can find a constant
unitary matrix $T_{s\times s}$, which makes $T^{\dag }h_{s\times
s}(q)T=-h_{s\times s}(q)$,\ $H_{s\times s}(q)$\ should possess one band with
an exactly zero orbital magnetic moment. We also give the $T$ matrix for Eq.
(\ref{Hk.p}), and its existence is originated from the RS constraints (See
Sec. IX of \cite{SM}$\ $for the detail).

\section{Material realizations of non-relativistic RS insulator.}

We now explore the possible materials to realize the non-relativistic 2D
M-RS quasiparticles, i.e., the low energy $\mathbf{k}\cdot \mathbf{p}$
Hamiltonian {bears} the same form as $\tilde{H}_{\pm }$ in Eq. (\ref{Hk.p}).
{To realize the $3\times 3$\ matrices $\tilde{H}_{\pm }$, we consider a}
layered material with three layers consisting of the same atoms. Furthermore
those specific atoms should comprise the identical planar structures for the
three layers in order for the coefficients of the diagonal term $%
q_{x}^{2}+q_{y}^{2}$ to be the same. The same coefficients also require
three identical atomic orbitals to be the basis of the low energy $\mathbf{k}%
\cdot \mathbf{p}$ Hamiltonian. The layered structure makes the $z$ direction
different from the $x$ and $y$ directions in the atomic plane. Thus there is
an energy splitting between $p_{z}$ and $p_{x,y}$ orbitals for example. Here
we take three $p_{z}$ orbitals as the basis set of the low energy $\mathbf{k}%
\cdot \mathbf{p}$ Hamiltonian. On the other hand, from the perspective of
symmetry, we firstly consider the allowable rotation symmetry operations
along the $z$ axis. The diagonal terms have the form of $q_{x}^{2}+q_{y}^{2}$%
, which requires the Hamiltonian at the RS point to own $C_{3},C_{4}$ or $%
C_{6}$\ rotation symmetry\cite{SM}. The form of the off-diagonal terms $%
\tilde{H}_{\pm ,12}$ and $\tilde{H}_{\pm ,13}$ ($q_{\mp } $) (namely the
coupling between different layers) rules out the possibilities of $C_{4}$
and $C_{6}$ rotation symmetry and the system belongs to either trigonal or
hexagonal crystal systems which possess $C_{3}$ rotation symmetry, while the
RS point should be $K$ point (or $K^{\prime }=-K $,$K=(1/3,1/3)$) located at
the corners of the hexagonal BZ\cite{SM}. The three layers should be of
equal distances for the coefficients of $q_{\mp }$ in $\tilde{H}_{\pm ,12}$
and $\tilde{H}_{\pm ,13}$ to be the same. Consequently the specific atoms in
each layer comprise a triangular lattice which is displayed as the
prototypical structure shown in Fig. S 1 of the SM \cite{SM}. The vanishing
terms $H_{\pm ,23}$ require that the hopping between the first and third
layers is negligible. As seen from Eq. (\ref{Hk.p}), $\tilde{H}_{+}$ or $%
\tilde{H}_{-}$, breaks the TR symmetry alone while they can be related to
each other in the TR related k points in the BZ through a TR transformation
to ensure the TR symmetry of the CMS. Thus we consider the non-magnetic
materials with the TR symmetry.

\begin{figure}[tbp]
\begin{center}
\includegraphics[width=7.5cm]{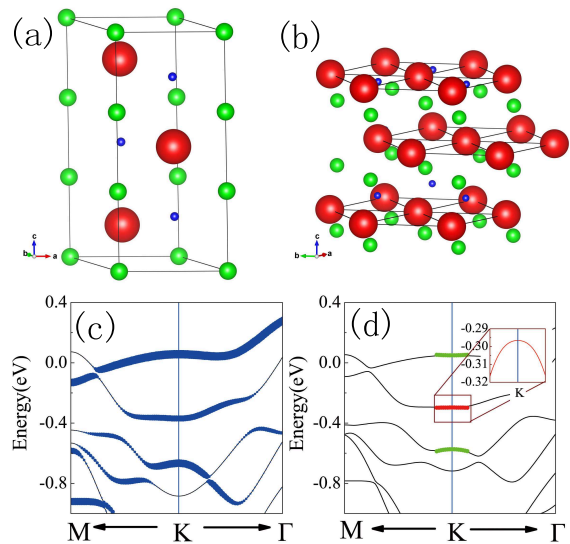}\\[0pt]
\end{center}
\caption{(a) Crystal structure of CaLiX (X=Pb, Ge and Si): the red ones
denote the X atoms, while the green ones and blue ones represent Ca atoms
and Li atoms, respectively. Three X atoms form three layers, and each layer
is a triangular lattice. (b) A 1-unit-cell-thick layer structure of CaLiX.
(c) The electronic structure of this thin film of CaLiPb: the blue fatted
curves are the projection bands for 6$p_{z}$ orbitals of Pb. (d) The band
structure of 1-unit-cell-thick layer of CaLiGe: the two green and one red
thick line are the energy dispersions obtained from our low energy effective
model Hamiltonian using the parameters in Table \protect\ref{parameters}.
The red curve represents the non-relativistic 2D M-RS band. }
\label{material}
\end{figure}

{To our knowledge, CaLiPb is a promising candidate.} CaLiPb crystallizes in
the hexagonal YbAgPb type structure with space group P$\overline{6}$m2 (No.
187) \cite{licapb}. The crystal structure of CaLiPb is shown as an example
in Fig. \ref{material}(a), in which the three Pb atoms are located at two
nonequivalent crystallographic sites: $2h(1/3,2/3,0.1533)$ and $%
1f(2/3,1/3,0.5)$, thus form three triangular lattices which have the same
structure as the prototypical structure. Considering the energy splittings
of the diagonal terms in the Hamiltonian of Eq. (\ref{Hk.p}), we exfoliate
CaLiPb into a 1-unit-cell-thick layer\textbf{\ }as shown in Fig. \ref%
{material}(b). The electronic structure of this thin film of CaLiPb is shown
in Fig. \ref{material}(c). The bands of Li ions and Ca ions are far away
from the Fermi level due the highly ionic property. At $K$ point, there are
considerable crystal splittings between Pb-$p_{z}$ and Pb-$p_{x}$/$p_{y}$
states as we expected. Hence, three bands near the Fermi level are mainly
contributed by Pb-$p_{z}$ states as shown in the fatted bands in Fig. \ref%
{material}(c). To avoid the large spin-orbit coupling in CaLiPb, we then
consider lighter elements like Ge and Si. Fig. \ref{material}(d) shows the
band structure of 2D CaLiGe which has the same lattice structure as shown in
Fig. \ref{material}(b). Similar to the electronic structure of CaLiPb, three
bands near the Fermi level are mainly contributed by Ge-$p_{z}$ states. %
By fitting the energy spectrum of the following Hamiltonian (\ref{h1}) with
that of the first-principles calculations, the parameters in the model
Hamiltonian (\ref{h1}) can be determined.

\begin{align}
\left(
\begin{array}{ccc}
\Delta_{1}(q_{x}^{2}+q_{y}^{2})+E_{1} & \Delta_{2}q_{\pm} & \Delta_{2}q_{\pm
} \\
\Delta_{2}q_{\mp} & \Delta_{1}(q_{x}^{2}+q_{y}^{2})+E_{2} & \Delta_{23} \\
\Delta_{2}q_{\mp} & \Delta_{23} & \Delta_{1}(q_{x}^{2}+q_{y}^{2})+E_{3}%
\end{array}
\right) .  \label{h1}
\end{align}

From the Fig. \ref{material}(d) we can see that in the vicinity of RS point%
\textbf{, }the model Hamiltonian Eq. (\ref{h1}) with the parameters given in
Table \ref{parameters} fits the first-principles results very well. The {%
non-relativistic} 2D M-RS mode is denoted by the red thick line in Fig. \ref%
{material}(d). As shown in Table \ref{parameters}, the parameter $\delta
_{23}$ is very small and can be negligible as expected. In addition to the
space group P$\overline{6}$m2, there are also other eight space groups
listed in Sec. X of the SM \cite{SM}. They can provide the possibilities of
the similar three-layer structure as shown in Fig. \ref{material}(b), thus
may also realize the non-relativistic 2D M-RS excitations.

\begin{table}[tbp]
\begin{tabular}{|l|l|l|l|l|l|}
\hline
$\Delta _{1}$(eV$\cdot $\AA $^{2}$) & $\Delta _{2}$(eV$\cdot $\AA ) & $E_{1}$%
(eV) & $E_{2}$(eV) & $E_{3}$(eV) & $\Delta _{23}$(eV) \\ \hline
-1.049 & -0.852 & -0.296 & 0.049 & -0.573 & -0.008 \\ \hline
\end{tabular}%
\caption{The parameters in the Hamiltonian (\protect\ref{h1}) after fitting
the energy bands by first principles calculation.}
\label{parameters}
\end{table}

\section{Discussions and Conclusions.}

We studied the counterpart of spin-$3/2$ M-RS particles in CMS. We proposed
a Hamiltonian, which implicitly contains the nontrivial constraints in RS\
equations and has the eigen-states exactly the same as the RS vector-spinor
wave functions. Based on the perturbation method in terms of $p/m$, we
successfully eliminated the non-local terms in the M-RS Hamiltonian, and
reduced the size of the Hamiltonian matrix for 2D {non-relativistic} M-RS
quasi fermion. We found that the orbital magnetic moment for the {%
non-relativistic} RS-band is zero while its Berry curvature and orbital
magnetization are finite. We also give a criterion, which can justify if one
Hamiltonian has a zero-orbital-moment band or not. The possible lattice
structures and materials were proposed. An appropriate choice of the free
parameters in our proposed Hamiltonian can separate the RS bands from the
non-RS bands. {Furthermore, this non-relativistic RS insulator possesses TR
symmetry with the TR transformation $T$ satisfying $T^2=1$, thus, it fits in
the AI class of the ten-fold way of the classification of topological
insulators\cite{10fold}. Though a topologically trivial phase is expected,
we believe in the non-relativistic RS insulator a finer classification
exists because of the non-trivial constraints.} In addition to the compounds
proposed in this work, we believe by tuning these parameters (namely $\bar{H}%
_{1t}$) one can realize M-RS excitations in many different CMS.

It is worth mentioning that other higher spin relativistic particles (i.e.,
spin larger than $3/2$) also need nontrivial constraints to project out the
redundant DOF\cite{Dirac1}. Thus the strategy we proposed for constructing
the Hamiltonian for the equation of motion with constraints can also be used
for the higher spin fermions and bosons such as the gravitons or any other
constrained systems, and their existences and properties are possible to be
explored in CMS.

{Another intriguing theoretical method in dealing with constrained 2D system
is the parton construction where the auxiliary gauge fields eliminate the
redundant DOF\cite{parton}. Our method has the advantage of being operable
while through the parton construction, we may be able to determine the
fractional statistics of the 2D RS fermion which could be entirely different
from the known ones because of the constraints.}

Our {paper} provides an opportunity to study the novel properties of RS
particles. Furthermore, in the presence of various types of interactions as
well as the external magnetic field, this new emergent relativistic
excitations in CMS may bring much more fruitful and intriguing phenomena and
these study are under way.


\acknowledgments FT acknowledges Shusheng Xu, Rui Wang and Kai Li for
helpful discussions. We thank Yi Zhang for drawing our attention to the
parton construction. This work was supported by the National Key R\&D
Program of China (No. 2017YFA0303203), NSFC under Grants No. 11525417 and
No. 11374137 (FT, YPD, XGW), No. 11474061 (XL, YY).


\begin{thebibliography}{99}
\bibitem{note} Here $j$ stands for either $S$ for $d=3$ or $S^{z}$ for $d=2$%
. Correspondingly, $\phi_{3/2}$ is of eight components or two-components
while $\phi_{1/2}$ and $\psi_{d+1}$ are of four or two components.
\end{thebibliography}
\end{document}


\title{Supplemental Material}
\author{Feng Tang$^{1,4}$}
\thanks{These two authors contribute equally.}
\author{Xi Luo$^{2}$}
\thanks{These two authors contribute equally.}
\author{Yongping Du$^{5,1,4}$}
\author{Yue Yu$^{3,4}$}
\author{Xiangang Wan$^{1,4}$}
\affiliation{${}^{1}$National Laboratory of Solid State Microstructures and School of
Physics, Nanjing University, Nanjing 210093, China}
\affiliation{${}^{2}$CAS Key Laboratory of Theoretical Physics, Institute of Theoretical
Physics, Chinese Academy of Sciences, Beijing 100190, China}
\affiliation{${}^{3}$State Key Laboratory of Surface Physics, Center for Field Theory and
Particle Physics, Department of Physics, Fudan University, Shanghai 200433,
China}
\affiliation{${}^{4}$Collaborative Innovation Center of Advanced Microstructures, Nanjing
210093, China}
\affiliation{${}^{5}$Department of Applied Physics, Nanjing University of Science and
Technology, Nanjing 210094, China}
\maketitle

\section{ Notations}

In quantum field theory, there are two kinds of notations for the
four-vector or three-vector in 3+1 or 2+1 dimensional space-time, the
imaginary time Euclidian vector $x_{\mu }=(\mathbf{x},it)$ \cite{lurie} or
the real time Minkowski vector $x^{\mu}=(t,\mathbf{x})$ \cite{qft}. The
Greek letters $\mu,\nu,\ldots$ take values from 1 to $d+1$ for the Euclidian
notation and from 0 to $d$ for the Minkowski notation. The Latin letters $%
i,j,\ldots$ take values from $1$ to $d$ to denote the spatial index while $%
d+1$ or 0 denotes the temporal index. In the Euclidian notation, we do not
need to distinguish between the superscript and subscript of a vector and
the metric tensor is Kronecker-delta function $\delta_{\mu\nu}$. In the
Minkowski notation a contracovariant vector $x^{\mu}$ is labeled by a
superscript while a covariant vector $x_{\mu}$ is labeled by a subscript $%
x_{\mu}=g_{\mu\nu}x^{\nu}$, where the metric tensor $g_{\mu\nu}=$diag$%
(1,-1,-1,-1)$. In this work, we adopt the Euclidian notation. The
vector-spinor which describes the Rarita-Schwinger (RS) field is denoted by $%
\psi _{\mu,\alpha}$ or $\psi_{\mu} $(omitting the spinor index $\alpha$).
The vector spinor in the Minkowski notation can be obtained by the following
substitution, $\psi_{i}\rightarrow\psi^{i}$ and $\psi_{d+1}\rightarrow
i\psi^{0}$. Any form like $A\cdot B$ means a summation over the vector
index, namely $\sum\limits_{\mu}A_{\mu}B_{\mu}$. We use the units that $%
c=\hbar=1$.

\section{ Gamma matrices}

We use ${\alpha_{i}}^{\prime}s$ and $\beta$ to express the $\gamma$
matrices: $\gamma_{i}=-i\beta\alpha_{i}$ and $\gamma_{d+1}=\beta$ \cite%
{lurie}. Dirac equation $(\gamma\cdot\partial+m)\psi=0$ can be rewritten in
a Hamiltonian form $i\partial_{t}\psi=H_{D}\psi$ where $H_{D}=\boldsymbol{%
\alpha }\cdot\mathbf{p}+\beta m$ \cite{lurie}. In order to satisfy the
relativistic energy-momentum relation, $\{\alpha_{i},\alpha_{j}\}=2%
\delta_{ij},\{\alpha_{i},\beta\}=0,\beta^{2}=1$ and then $%
\{\gamma_{\mu},\gamma_{\nu}\}=\delta_{\mu\nu}$. Hence, $\gamma_{\mu}$ are $%
4\times4$ matrices for $d=3$ while $2\times2$ matrices for $d=2$.

In three dimensional (3D) systems, we use the Dirac representation for the
spinor where, $\alpha _{i}=\left(
\begin{array}{cc}
0 & \sigma _{i} \\
\sigma _{i} & 0%
\end{array}%
\right) ,$ and $\beta =\left(
\begin{array}{cc}
\sigma _{0} & 0 \\
0 & -\sigma _{0}%
\end{array}%
\right) .$ The Pauli matrices are given by $\sigma _{1}=\left(
\begin{array}{cc}
0 & 1 \\
1 & 0%
\end{array}%
\right) $,$\sigma _{2}=\left(
\begin{array}{cc}
0 & -i \\
i & 0%
\end{array}%
\right) $, and $\sigma _{3}=\left(
\begin{array}{cc}
1 & 0 \\
0 & -1%
\end{array}%
\right) $. $\sigma _{0}=I_{2\times 2}$ is the identity matrix. In this
representation, for the rest coordinate frame (i.e., $\mathbf{p}=(0,0,0)$),
the Dirac Hamiltonian is already diagonalized: diag$(m,m,-m,-m)$, and one
can denote the spinor index as $(\xi \sigma )$ where $\xi =\pm $ and $\sigma
=\uparrow ,\downarrow $. In the Weyl representation, the representation of
the Lorentz group is $(\frac{1}{2},0)\oplus (0,\frac{1}{2})$ \cite{qft}. The
spinor index is denoted by $(\zeta \sigma )$ where $\zeta =L,R$ and $\sigma
=\uparrow ,\downarrow $. The spinors in the Dirac representation are related
to the spinors in the Weyl representation by $\psi _{+,\sigma }=\frac{1}{%
\sqrt{2}}(\psi _{L,\sigma }-\psi _{R,\sigma }),$ $\psi _{-,\sigma }=\frac{1}{%
\sqrt{2}}(\psi _{L,\sigma }+\psi _{R,\sigma })$. We use the Dirac
representation in this work. In two dimensional (2D) systems (located in the $xy$ plane) where $p_{3}$
is set to be zero, it is convenient to use three Pauli matrices to represent
$\alpha _{1},\alpha _{2}$ and $\beta $, and we can choose $\alpha _{1}=\sigma _{1}$, $\alpha _{2}=\sigma _{2}$ and $%
\beta =\sigma _{3}$. We will discuss the dimensional reduction from $d=3$ to
$d=2$ in Sec. V.



\section{3D RS Hamiltonians}

In this section, we derive the 3D RS Hamiltonians. We define $\mathbf{\Sigma}%
^{\frac{3}{2}}=\frac{2}{3}\mathbf{J}^{\frac{3}{2}}$ where $\mathbf{J}^{\frac{%
3}{2}}$ are $4\times4$ vector matrices of the angular momentum operator in
the representation $j=\frac{3}{2}$ of SO(3) group with the following basis
set:
\begin{equation*}
|\frac{3}{2},\frac {3}{2}>,|\frac{3}{2},-\frac{3}{2}>,|\frac{3}{2},\frac{1}{2%
}>,|\frac{3}{2},-\frac{1}{2}>
\end{equation*}%
. Namely,
\begin{eqnarray}
J^{\frac{3}{2}}_{1}&=&\left(
\begin{array}{cccc}
0 & 0 & \frac{\sqrt3}{2} & 0 \\
0 & 0 & 0 & \frac{\sqrt3}{2} \\
\frac{\sqrt3}{2} & 0 & 0 & 1 \\
0 & \frac{\sqrt3}{2} & 1 & 0%
\end{array}
\right),  \notag \\
J^{\frac{3}{2}}_{2}&=&\left(
\begin{array}{cccc}
0 & 0 & -i\frac{\sqrt3}{2} & 0 \\
0 & 0 & 0 & i\frac{\sqrt3}{2} \\
i\frac{\sqrt3}{2} & 0 & 0 & -i \\
0 & -i\frac{\sqrt3}{2} & i & 0%
\end{array}
\right) ,  \notag \\
J^{\frac{3}{2}}_{3}&=&\left(
\begin{array}{cccc}
\frac{3}{2} & 0 & 0 & 0 \\
0 & -\frac{3}{2} & 0 & 0 \\
0 & 0 & \frac{1}{2} & 0 \\
0 & 0 & 0 & -\frac{1}{2}%
\end{array}
\right) .  \notag
\end{eqnarray}
The space-like spinors $\psi_{1}$, $\psi_{2}$, $\psi_{3}$ (or $%
\psi_{x},\psi_{y},\psi_{z}$) are a product representation $1\otimes \frac{1}{%
2}$ of SO(3) group. They are related to the $j=\frac{3}{2}$ and $j=\frac{1}{2%
}$ irreducible representations by
\begin{widetext}
\begin{equation}\label{3dutm}
\left(\begin{array}{c}
\psi_{1,\xi,\uparrow}\\ \psi_{1,\xi,\downarrow}\\  \psi_{2,\xi,\uparrow}\\ \psi_{2,\xi,\downarrow}\\  \psi_{3,\xi,\uparrow}\\ \psi_{3,\xi,\downarrow}\\
\end{array}
\right)=
\left(\begin{array}{cccccc}
-\frac{1}{\sqrt2}& 0& 0& \frac{1}{\sqrt6}& 0& -\frac{1}{\sqrt3}\\
0& \frac{1}{\sqrt2}& -\frac{1}{\sqrt6}& 0& -\frac{1}{\sqrt3}& 0\\
-\frac{i}{\sqrt2}& 0& 0& -\frac{i}{\sqrt6}& 0& \frac{i}{\sqrt3}\\
0& -\frac{i}{\sqrt2}& -\frac{i}{\sqrt6}& 0& -\frac{i}{\sqrt3}& 0\\
0& 0& \sqrt{\frac{2}{3}}& 0& -\frac{1}{\sqrt3}& 0\\
0& 0& 0& \sqrt{\frac{2}{3}}& 0& \frac{1}{\sqrt3}
\end{array}\right)
\left(\begin{array}{c}
\phi_{\xi,\frac{3}{2},-\frac{3}{2}}\\ \phi_{\xi,\frac{3}{2},-\frac{3}{2}}\\  \phi_{\xi,\frac{3}{2},\frac{1}{2}}\\ \phi_{\xi,\frac{3}{2},-\frac{1}{2}}\\  \phi_{\xi,\frac{1}{2},\frac{1}{2}}\\ \phi_{\xi,\frac{1}{2},-\frac{1}{2}}\\
\end{array}\right)\equiv R\left(\begin{array}{c}
\phi_{\xi,\frac{3}{2},-\frac{3}{2}}\\ \phi_{\xi,\frac{3}{2},-\frac{3}{2}}\\  \phi_{\xi,\frac{3}{2},\frac{1}{2}}\\ \phi_{\xi,\frac{3}{2},-\frac{1}{2}}\\  \phi_{\xi,\frac{1}{2},\frac{1}{2}}\\ \phi_{\xi,\frac{1}{2},-\frac{1}{2}}\\
\end{array}\right).
\end{equation}
\end{widetext}
In $\phi_{\xi,\frac{3}{2},\frac{3}{2}}$, the first $\frac{3}{2}$ represents $%
j$ and the second represents $j_{z}$. We use $\phi_{\xi,\frac{3}{2}}$ to
represent $(\phi_{\xi,\frac{3}{2},\frac{3}{2}},\phi_{\xi,\frac{3}{2},-\frac{3%
}{2}},\phi_{\xi,\frac{3}{2},\frac{1}{2}},\phi_{\xi,\frac{3}{2},-\frac{1}{2}%
})^{T}$, and $\phi_{\xi,\frac{1}{2}}$ to represent $(\phi_{\xi,\frac{1}{2},%
\frac{1}{2}},\phi_{\xi,\frac{1}{2},-\frac{1}{2}})^T$. According to $%
(\gamma\cdot\partial+m)\psi_{i}=0$ or $i\partial_{t}\psi_{i}=H_{D}\psi_{i}$
in the RS equations, we get the equations of motion of $\phi_{\xi,\frac{3}{2}%
}$ and $\phi_{\xi,\frac{1}{2}}$:

\begin{equation}
i\partial_{t}\phi_{\xi,\frac{3}{2}}=h_{33}^{\prime}(\mathbf{p})\phi_{\bar{%
\xi },\frac{3}{2}}+h_{31}^{\prime}(\mathbf{p})\phi_{\bar{\xi},\frac{1}{2}%
}+\xi m\phi_{\xi,\frac{3}{2}},  \label{3drs3}
\end{equation}%
\begin{equation}
i\partial_{t}\phi_{\xi,\frac{1}{2}}=h_{13}^{\prime}(\mathbf{p})\phi_{\bar{%
\xi },\frac{3}{2}}+h_{11}^{\prime}(\mathbf{p})\phi_{\bar{\xi},\frac{1}{2}%
}+\xi m\phi_{\xi,\frac{1}{2}},  \label{3drs1}
\end{equation}
where $h^{\prime}=R^{\dag}(I_{3\times3}\otimes\boldsymbol{\sigma }\cdot%
\mathbf{p})R$. Namely,
\begin{equation}
h^{\prime}(\mathbf{p})=\left(
\begin{array}{cc}
h_{33}^{\prime} & h_{31}^{\prime} \\
h_{13}^{\prime} & h_{11}^{\prime}%
\end{array}
\right) ,  \label{h}
\end{equation}
where $h_{33}^{\prime}=\mathbf{p}\cdot\mathbf{\Sigma}^{\frac{3}{2}}$, $%
h_{11}^{\prime}=-\frac{1}{3}\mathbf{p}\cdot\boldsymbol{\sigma}$, and $%
h_{31}^{\prime}={h^{\prime}}_{13}^{\dag}=\left(
\begin{array}{cc}
\sqrt{\frac{2}{3}}p_{-} & 0 \\
0 & -\sqrt{\frac{2}{3}}p_{+} \\
-\frac{2\sqrt{2}}{3}p_{3} & \frac{\sqrt{2}}{3}p_{-} \\
-\frac{\sqrt{2}}{3}p_{+} & -\frac{2\sqrt{2}}{3}p_{3}%
\end{array}
\right) ,$ with $p_{-}=p_{+}^{*}=p_{1}-ip_{2}.$\newline

Simultaneously the constraints $\gamma \cdot \psi =0$ in the RS equations
become
\begin{equation}
\psi _{4,\xi }=-i\sqrt{3}\phi _{\bar{\xi},\frac{1}{2}}.  \label{3drt1}
\end{equation}%
Substituting this equation into $(\gamma \cdot \partial +m)\psi _{4}=0$ or $%
i\partial _{t}\psi _{4}=H_{D}\psi _{4}$, we get
\begin{equation}
i\partial _{t}\phi _{\xi ,\frac{1}{2}}=\boldsymbol{\sigma }\cdot \mathbf{p}%
\phi _{\bar{\xi},\frac{1}{2}}-\xi m\phi _{\xi ,\frac{1}{2}}.  \label{3drs1a}
\end{equation}%
As we will show later, Eq. (\ref{3drs3}) and Eq. (\ref{3drs1}) correspond to
Eq. (5) and Eq. (6) in the main text, respectively; Eq. (\ref{3drt1}) and
Eq. (\ref{3drs1a}) correspond to Eq. (8) and Eq. (9) in the main text,
respectively. Combining Eq. (\ref{3drs1a}) and Eq. (\ref{3drs1}), we get the
relation between $\phi _{\xi ,\frac{3}{2}}$ and $\phi _{\xi ,\frac{1}{2}}$:
\begin{equation}
h_{13}^{\prime }\phi _{\xi ,\frac{3}{2}}+4h_{11}^{\prime }\phi _{\xi ,\frac{1%
}{2}}=2m\xi \phi _{\bar{\xi},\frac{1}{2}},  \label{3dr13-0}
\end{equation}%
from which we have
\begin{equation}
\phi _{\xi ,\frac{1}{2}}=(m^{2}+4{h^{\prime }}_{11}^{2})^{-1}(-{h^{\prime }}%
_{11}{h^{\prime }}_{13}\phi _{\xi ,\frac{3}{2}}+\xi \frac{m}{2}{h^{\prime }}%
_{13}\phi _{\bar{\xi},\frac{3}{2}}),  \label{3dr13}
\end{equation}%
which corresponds to Eq. (11) in the main text. It is clear that $\phi _{\xi
,\frac{1}{2}}$\ and $\psi _{4,\xi }$\ are indeed redundant degrees of freedom (DOF) and can be
obtained from genuine DOF $\phi _{\xi ,\frac{3}{2}}$ with Eq. (\ref{3drt1})
and Eq. (\ref{3dr13}). By substituting Eq. (\ref{3dr13}) into Eq. (\ref%
{3drs3}), we find that the Hamiltonian of $\phi _{\xi ,\frac{3}{2}}$,
\begin{widetext}\begin{equation}\label{3drs3H-0}
i\partial_t\phi_{\xi,\frac{3}{2}}=(h'_{33}-(m^2+4{h'}_{11}^2)^{-1}{h'}_{31}{h'}_{11}{h'}_{13})\phi_{\bar{\xi},\frac{3}{2}}+\xi m(1+\frac{1}{2}(m^2+4{h'}_{11}^2)^{-1}{h'}_{31}{h'}_{13}) \phi_{\xi,\frac{3}{2}}.
\end{equation}
\end{widetext}Notice that ${h^{\prime }}_{11}^{2}=\frac{1}{9}p^{2},-{%
h^{\prime }}_{31}{h^{\prime }}_{11}{h^{\prime }}_{13}=(\mathbf{\Sigma }^{%
\frac{3}{2}}\cdot \mathbf{p})(p^{2}-(\mathbf{\Sigma }^{\frac{3}{2}}\cdot
\mathbf{p})^{2})$, and ${h^{\prime }}_{31}{h^{\prime }}_{13}=p^{2}-(\mathbf{%
\Sigma }^{\frac{3}{2}}\cdot \mathbf{p})^{2}$. We can then write the above
equation in a more compact form:
\begin{widetext}\begin{eqnarray}\label{3drs3H}
i\frac{\partial}{\partial t}\phi_{\xi,\frac{3}{2}}=
{\bf \Sigma}^{\frac{3}{2}}\cdot{\bf p}[1+\frac{p^2-({\bf \Sigma}^{\frac{3}{2}}\cdot{\bf p})^2}{\frac{4}{9}p^2+m^2}]\phi_{\bar{\xi},\frac{3}{2}}+\xi m[1+\frac{1}{2}\frac{p^2-({\bf\Sigma}^{\frac{3}{2}}\cdot{\bf p})^2}{\frac{4}{9}p^2+m^2}]\phi_{\xi,\frac{3}{2}}.
\end{eqnarray}
\end{widetext}
If we represent $(\phi _{+,\frac{3}{2}},\phi _{-,\frac{3}{2}%
})^{T}$ by $\phi _{3/2}$, and $(\phi _{+,\frac{1}{2}},\phi _{-,\frac{1}{2}%
})^{T}$ by $\phi _{1/2}$, (\ref{3drt1}) and (\ref{3dr13}) can be written in
the following form
\begin{equation}
\psi _{4}=-i\sqrt{3}\sigma _{1}\otimes \sigma _{0}\phi _{1}\equiv R_{t1}\phi
_{1/2},  \label{3drt1-1}
\end{equation}%
\begin{equation}
\phi _{1/2}=\frac{1}{\frac{4}{9}p^{2}+m^{2}}(-\sigma _{0}\otimes {h^{\prime }}%
_{11}{h^{\prime }}_{13}+\frac{m}{2}i\sigma _{2}\otimes {h^{\prime }}%
_{13})\phi _{3/2}\equiv R_{13}\phi _{3/2}.  \label{3dr13-1}
\end{equation}%
Eq. (\ref{3drs3}), Eq. (\ref{3drs1}) and Eq. (\ref{3drs1a}) can be written
in the following compact form:
\begin{align}
i\partial _{t}\phi _{3/2}& =h_{33}\phi _{3}+h_{31}\phi _{1/2},  \label{3drs3-1}
\\
i\partial _{t}\phi _{1/2}& =h_{13}\phi _{3}+h_{11}\phi _{1/2},  \label{3drs1-1}
\\
i\partial _{t}\phi _{1/2}& =H_{1/2}\phi _{1/2},  \label{3drs1a-1}
\end{align}%
where, $h_{33}=\left(
\begin{array}{cc}
m & h_{33}^{\prime } \\
h_{33}^{\prime } & -m%
\end{array}%
\right) $, $h_{31}=\left(
\begin{array}{cc}
0 & h_{31}^{\prime } \\
h_{31}^{\prime } & 0%
\end{array}%
\right) $,$h_{11}=\left(
\begin{array}{cc}
m & h_{11}^{\prime } \\
h_{11}^{\prime } & -m%
\end{array}%
\right) $ and $H_{1/2}=\boldsymbol{\alpha }\cdot \mathbf{p}-\beta
m=R_{t1}^{-1}H_{t}R_{t1}.$ Eq. (\ref{3drs3H}) can also be written as,

\begin{equation}  \label{3drs3H-a}
i\partial_{t}\phi_{3/2}=H_{3/2}\phi_{3/2},
\end{equation}
where $H_{3/2}=\sigma_{1}\otimes(\mathbf{\Sigma}^{\frac{3}{2}}\cdot \mathbf{p}%
)[1+\frac{p^{2}-(\mathbf{\Sigma}^{\frac{3}{2}}\cdot\mathbf{p})^{2}}{\frac{4}{%
9}p^{2}+m^{2}}] +m\sigma_{3}\otimes[1+\frac{1}{2}\frac {p^{2}-(\mathbf{\Sigma%
}^{\frac{3}{2}}\cdot\mathbf{p})^{2}}{\frac{4}{9}p^{2}+m^{2}}]$.

\section{2D RS Hamiltonians}

The situation for the 2D RS systems is similar to that in the 3D RS systems.
In the 2D systems, the typical elements of the Lorentz group would be just
the boosts along the $x$- and $y$-directions as well as the pure rotations
around the $z$-axis. The pure rotation group is then SO(2). To characterize
the angular momentum we merely need one number $j_{z}$ instead of $(j,j_{z})$
in the 3D systems. In the 2D systems, the vector spinor field would be
explicitly written as follows for the sake of clarity: $\psi _{1}(\mathbf{x}%
,t)=\left(
\begin{array}{c}
\psi _{1}(\mathbf{x},t,\uparrow ) \\
\psi _{1}(\mathbf{x},t,\downarrow )%
\end{array}%
\right) $, $\psi _{2}(\mathbf{x},t)=\left(
\begin{array}{c}
\psi _{2}(\mathbf{x},t,\uparrow ) \\
\psi _{2}(\mathbf{x},t,\downarrow )%
\end{array}%
\right) $, $\psi _{3}(\mathbf{x},t)=\left(
\begin{array}{c}
\psi _{3}(\mathbf{x},t,\uparrow ) \\
\psi _{3}(\mathbf{x},t,\downarrow )%
\end{array}%
\right) ,$ where $\psi _{1},\psi _{2}$ (or $\psi _{x},\psi _{y}$) are
space-like spinors while $\psi _{3}$ is a time-like spinor. In the momentum
representation, the RS equations would be written as follows:
\begin{align}
i\partial _{t}\psi _{1}(\mathbf{p},t)& =(\boldsymbol{\sigma }\cdot \mathbf{p}%
+m\sigma _{3})\psi _{1}(\mathbf{p},t),  \label{dirac1} \\
i\partial _{t}\psi _{2}(\mathbf{p},t)& =(\boldsymbol{\sigma }\cdot \mathbf{p}%
+m\sigma _{3})\psi _{2}(\mathbf{p},t),  \label{dirac2} \\
i\partial _{t}\psi _{3}(\mathbf{p},t)& =(\boldsymbol{\sigma }\cdot \mathbf{p}%
+m\sigma _{3})\psi _{3}(\mathbf{p},t),  \label{dirac3}
\end{align}%
with the constraint: $\psi _{3}(\mathbf{p},t)=i\sigma _{i}\psi _{i}(\mathbf{p%
},t)$.

We would make the following unitary transformation in order to extract the
genuine DOF which are in the space of $j_{z}=\pm\frac{3}{2}$ representations
of SO(2):
\begin{widetext}
\begin{equation}\label{2dutm}
\left(\begin{array}{c}\psi_{1,\uparrow}\\
\psi_{1,\downarrow}\\
\psi_{2,\uparrow}\\
\psi_{2,\downarrow}\\
\end{array}\right)=\left(\begin{array}{cccc}
-\frac{1}{\sqrt2}& 0& 0& -\frac{1}{\sqrt2}\\
0& \frac{1}{\sqrt2}& -\frac{1}{\sqrt2}& 0\\
-\frac{i}{\sqrt2}& 0& 0& \frac{i}{\sqrt2}\\
0& -\frac{i}{\sqrt2}& -\frac{i}{\sqrt2}& 0
\end{array}\right) \left(\begin{array}{c}\phi_{j_z=\frac{3}{2}}\\
\phi_{j_z=-\frac{3}{2}}\\
\phi_{j_z=\frac{1}{2}}\\
\phi_{j_z=-\frac{1}{2}}\\
\end{array}\right),
\end{equation}
\end{widetext}
where $\phi_{j_{z}=\pm\frac{3}{2}, \pm\frac{1}{2}}$ transform
as $j_{z}$ representation under SO(2). We can easily transform (\ref{dirac1}%
) and (\ref{dirac2}) into the above new representation. Denoting $(\phi_{%
j_z=\frac{3}{2}},\phi_{j_z=-\frac{3}{2}})^{T}$ as $\phi_{3/2}$ and $(\phi _{j_z=\frac{1}{2}%
},\phi_{j_z=-\frac{1}{2}})^{T}$ as $\phi_{1/2}$:
\begin{equation}  \label{2drs3}
i\partial_{t}\phi_{3/2}=h_{33}\phi_{3/2}+h_{31}\phi_{1/2},
\end{equation}
\begin{equation}  \label{2drs1}
i\partial_{t}\phi_{1/2}=h_{13}\phi_{3/2}+h_{11}\phi_{1/2},
\end{equation}
where
\begin{equation}  \label{2drshhh}
h_{33}=h_{11}=-\sigma_{3}m,h_{31}=h_{13}^{\dag}=\mathrm{diag}(p_{-},-p_{+}),
\end{equation}
and the constraint (\ref{dirac3}) would be,
\begin{equation}  \label{2drt1}
\psi_{3}=-i\sqrt2\phi_{1/2}\equiv R_{t1}\phi_{1/2}.
\end{equation}
As (\ref{dirac3}) describes the equation of motion of $\psi_{3}$, its
Hamiltonian $H_{t}$ is given by
\begin{equation}  \label{2drsht}
H_{t}=\boldsymbol{\alpha}\cdot\mathbf{p}+\sigma_{3}m.
\end{equation}

Substituting (\ref{2drt1}) into (\ref{dirac3}) we will get the Hamiltonian
of $\phi_{1}$,
\begin{equation}  \label{2drs1a}
i\partial_{t}\phi_{1/2}=H_{1/2}\phi_{1/2},
\end{equation}
where $H_{1/2}=R_{t1}^{-1}H_{t}R_{t1}=\boldsymbol{\sigma}\cdot\mathbf{p}%
+\sigma _{3}m$. Combining (\ref{2drs1}) and (\ref{2drs1a}), we get,
\begin{equation}  \label{2drs13}
\phi_{1/2}= \frac{1}{4m^{2}+p^{2}} \left(
\begin{array}{cc}
2mp_{+} & -p_{-}^{2} \\
p_{+}^{2} & 2mp_{-}%
\end{array}
\right) \phi_{3/2}\equiv R_{13}\phi_{3/2}.
\end{equation}
Substituting (\ref{2drs13}) into (\ref{2drs3}), we have
\begin{equation}  \label{2drs3h}
i\partial_{t}\phi_{3/2}=H_{3/2}\phi_{3/2},
\end{equation}
where $H_{3/2}=\left(
\begin{array}{cc}
m+\frac{2mp^{2}}{4m^{2}+p^{2}} & -\frac{p_{-}^{3}}{4m^{2}+p^{2}} \\
-\frac{p_{+}^{3}}{4m^{2}+p^{2}} & -(m+\frac{2mp^{2}}{4m^{2}+p^{2}})%
\end{array}
\right). $ Notice that when setting $p_{3}=0$ and $\psi_{3}=(0,0,0,0)^{T}$
in three dimensions, the results in the 3D RS systems will turn into those
of the 2D RS systems.

\section{From $d=3$ to $d=2$}

For the Dirac field, when the spatial dimension is reduced from three to two
and the 2D system is located in the $xy$ plane, we can set $p_{3}=0$. We use
the superscripts $3D$ or $2D$ to distinguish between the quantities in 3D or
2D. It is easy to find that $H_{D}^{3D}$ can be written in a
block-diagonalized form when we write the wave function as $(\psi
_{+,\uparrow }^{3D},\psi _{-,\downarrow }^{3D},\psi _{-,\uparrow }^{3D},\psi
_{+,\downarrow }^{3D})^{T}$:
\begin{equation}
H_{D}^{3D}=\left(
\begin{array}{cccc}
m & p_{-} & 0 & 0 \\
p_{+} & -m & 0 & 0 \\
0 & 0 & -m & p_{-} \\
0 & 0 & p_{+} & m%
\end{array}%
\right) .  \label{3Dto2D-Dirac}
\end{equation}%
Then the 3D Dirac equation will be reduced to two 2D Dirac equations with
masses of opposite signs, i.e. $\pm m$. If we define $\psi _{\uparrow
}^{2D,m}=\psi _{+,\uparrow }^{3D},\psi _{\downarrow }^{2D,m}=\psi
_{-,\downarrow }^{3D}$ and $\psi _{\uparrow }^{2D,-m}=\psi _{-,\uparrow
}^{3D},\psi _{\downarrow }^{2D,-m}=\psi _{+,\downarrow }^{3D}$, then the two
2D Dirac equations would be, $i\partial _{t}\psi ^{2D,\pm m}=(\sigma
_{1}p_{1}+\sigma _{2}p_{2}\pm \sigma _{3}m)\psi ^{2D,\pm m}$.\newline
As for the RS field, we will show how the 3D RS vector spinor can be reduced
to two 2D RS vector spinors by setting $p_{3}=0$ and $\psi
_{3}^{3D}=(0,0,0,0)^{T}$. Analogously as the Dirac field, we expect that by
setting $\psi _{i,\uparrow }^{2D,m}=\psi _{i,+,\uparrow }^{3D},\psi
_{i,\downarrow }^{2D,m}=\psi _{i,-,\downarrow }^{3D}$ and $\psi _{i,\uparrow
}^{2D,-m}=\psi _{i,-,\uparrow }^{3D},\psi _{i,\downarrow }^{2D,-m}=\psi
_{i,+,\downarrow }^{3D}$ with $i=1,2$, the Dirac equations for $\psi
_{i}^{3D}$ will change to two decoupled 2D Dirac equations with masses of
opposite signs, i.e., $i\partial _{t}\psi _{i}^{2D,\pm m}=(\sigma
_{1}p_{1}+\sigma _{2}p_{2}\pm \sigma _{3}m)\psi _{i}^{2D,\pm m}$. For the
constraint, $\gamma ^{3D}\cdot \psi ^{3D}=0$, i.e., $\psi _{4}^{3D}=i\alpha
_{1}^{3D}\psi _{1}^{3D}+i\alpha _{2}^{3D}\psi _{2}^{3D}$, then $\left(
\begin{array}{c}
\psi _{4,+,\uparrow } \\
\psi _{4,+,\downarrow } \\
\psi _{4,-,\uparrow } \\
\psi _{4,-,\downarrow }%
\end{array}%
\right) =i\left(
\begin{array}{c}
\psi _{1,-,\downarrow }-i\psi _{2,-,\downarrow } \\
\psi _{1,-,\uparrow }+i\psi _{2,-,\uparrow } \\
\psi _{1,+,\downarrow }-i\psi _{2,+,\downarrow } \\
\psi _{1,+,\uparrow }+i\psi _{2,+,\uparrow }%
\end{array}%
\right) $. We can find that, $\left(
\begin{array}{c}
\psi _{4,+,\uparrow }^{3D} \\
\psi _{4,-,\downarrow }^{3D}%
\end{array}%
\right) =i\left(
\begin{array}{c}
\psi _{1,-,\downarrow }^{3D}-i\psi _{2,-,\downarrow }^{3D} \\
\psi _{1,+,\uparrow }^{3D}+i\psi _{2,+,\uparrow }^{3D}%
\end{array}%
\right) $, and $\left(
\begin{array}{c}
\psi _{4,-,\uparrow }^{3D} \\
\psi _{4,+,\downarrow }^{3D}%
\end{array}%
\right) =i\left(
\begin{array}{c}
\psi _{1,+,\downarrow }^{3D}-i\psi _{2,+,\downarrow }^{3D} \\
\psi _{1,-,\uparrow }^{3D}+i\psi _{2,-,\uparrow }^{3D}%
\end{array}%
\right) $. By setting $\psi _{3,\uparrow }^{2D,m}=\psi _{4,+,\uparrow
}^{3D},\psi _{3,\downarrow }^{2D,m}=\psi _{4,-,\downarrow }^{3D}$, and $\psi
_{3,\uparrow }^{2D,-m}=\psi _{4,-,\uparrow }^{3D},\psi _{3,\downarrow
}^{2D,-m}=\psi _{4,+,\downarrow }^{3D}$, we can find that for both $\psi
_{3}^{2D,m}$ and $\psi _{3}^{2D,-m}$, they satisfy $\psi _{3}^{2D,\pm
m}=i\sigma _{1}\psi _{1}^{2D,\pm m}+i\sigma _{2}\psi _{2}^{2D,\pm m}$, i.e.,
$\gamma ^{2D}\cdot \psi ^{2D,\pm m}=0$. Furthermore the time-like spinors in
2D: $\psi _{3}^{2D,\pm m}$ also satisfy the 2D Dirac equations: $i\partial
_{t}\psi _{3}^{2D,\pm m}=(\sigma _{1}p_{1}+\sigma _{2}p_{2}\pm \sigma
_{3}m)\psi _{3}^{2D,\pm m}$.\newline
In summary, the 3D RS equations are reduced to two sets of 2D RS equations
with masses $\pm m$: $i\partial _{t}\psi _{\mu }^{2D,\pm m}=(\sigma
_{1}p_{1}+\sigma _{2}p_{2}\pm \sigma _{3}m)\psi _{\mu }^{2D,\pm m}$ with a
nontrivial constraint $\gamma ^{2D}\cdot \psi ^{2D,\pm m}=0$ after setting $%
p_{3}=0$ and $\psi _{3}^{3D}=(0,0,0,0)^{T}$. According to the
transformations in (\ref{3dutm}) and (\ref{2dutm}) and using the above
relations between the vector spinors in 3D and 2D, it is easy to get
(keeping in mind that $\psi _{3}^{3D}=(0,0,0,0)^{T}$),
\begin{eqnarray}
&&\phi _{\xi ,\frac{3}{2},\frac{3}{2}}^{3D}=\phi _{\frac{3}{2}}^{2D,\xi m},
\\
&&\phi _{\xi ,\frac{3}{2},-\frac{3}{2}}^{3D}=\phi _{-\frac{3}{2}}^{2D,\bar{%
\xi}m}, \\
&&\phi _{\xi ,\frac{3}{2},\frac{1}{2}}^{3D}=\frac{1}{\sqrt{2}}\phi _{\xi ,%
\frac{1}{2},\frac{1}{2}}^{3D}=\frac{1}{\sqrt{3}}\phi _{\frac{1}{2}}^{2D,\bar{%
\xi}m} \\
&&\phi _{\xi ,\frac{3}{2},-\frac{1}{2}}^{3D}=-\frac{1}{\sqrt{2}}\phi _{\xi ,%
\frac{1}{2},-\frac{1}{2}}^{3D}=-\frac{1}{\sqrt{3}}\phi _{-\frac{1}{2}%
}^{2D,\xi m}.
\end{eqnarray}%
Because $\phi _{\xi ,\frac{3}{2},\pm \frac{1}{2}}^{3D}$ and $\phi _{\xi ,%
\frac{1}{2},\pm \frac{1}{2}}^{3D}$ are the same representations for SO(2)
group, they differ just by a constant. Using the above equations, we have
checked that the 3D results in Sec. III will recover those of 2D in Sec. IV.

\section{The selection of $\bar{H}_{1t}$}

In addition to the RS\ modes, the Hamiltonian in condensed matter systems
(CMS) (i.e., the $\bar{H}$\ in the Eq. (18) in the main text) contains the
other non-RS modes. Since we are only interested in the dispersion of the RS
modes, as shown in the main text, the only requirement for $\bar{H}_{1t}$\
is $R_{t1}^{\dag }\bar{H}_{1t}^{\dag }+\bar{H}_{1t}R_{t1}=0$, which cannot
fully determine the expression of $\bar{H}_{1t}$. For $d=2$, $R_{t1}=-i\sqrt{%
2}$, thus $\bar{H}_{1t}^{\dag }=\bar{H}_{1t}$, and $\bar{H}_{1t}$ has the
following form,
\begin{equation}
\bar{H}_{1t}=\left(
\begin{array}{cc}
a_{1} & b_{-} \\
b_{+} & a_{2}%
\end{array}%
\right) ,  \label{H1t}
\end{equation}%
where $b_{\pm}=b_{1}\pm ib_{2}$ and $a_{1},a_{2},b_{1},b_{2}$ are
independent real parameters. We plot the band spectrum with $\bar{H}%
_{1t}=m(\sigma _{1}+\sigma _{2})$ (i.e., $a_{1}=a_{2}=0,b_{1}=b_{2}=m$) in
Fig. 1 of the main text. It is clear that the RS modes are separated from
the non-RS\ modes. We also plot the bands for another selection of $%
a_{1}=a_{2}=0,b_{1}=b_{2}=2m$. It is clear that with changing $\bar{H}_{1t}$
by the restriction of $R_{t1}^{\dag }\bar{H}_{1t}^{\dag }+\bar{H}%
_{1t}R_{t1}=0$, the RS\ modes will not be modified, i.e., they are protected
from such a change, while the non-RS modes will be modified. And once $\bar{H%
}_{1t}\neq
\begin{pmatrix}
0 & 0 \\
0 & 0%
\end{pmatrix}%
$, the non-RS and RS\ modes are separated.

For $d=3$, $\bar{H}_{1t}^{\dag}=(\sigma_{1}\otimes\sigma_{0})\bar{H}%
_{1t}(\sigma_{1}\otimes\sigma_{0})$, then, if we write $\bar{H}_{1t}$ in a
blocking form,
\begin{equation}  \label{3dH1t}
\bar{H}_{1t}=\left(
\begin{array}{cc}
\bar{H}_{1t,++} & \bar{H}_{1t,+-} \\
\bar{H}_{1t,-+} & \bar{H}_{1t,--}%
\end{array}
\right) ,
\end{equation}
where $\bar{H}_{1t,\xi\xi^{\prime}}$($\xi,\xi^{\prime}=+/-$) is a $2\times2$
matrix, we will find $\bar{H}_{1t,--}=\bar{H}_{1t,++}^{\dag}$, $\bar {H}%
_{1t,+-}=\bar{H}_{1t,+-}^{\dag}$ and $\bar{H}_{1t,-+}=\bar{H}_{1t,-+}^{\dag}$%
. We then find that,

\begin{align}
\bar{H}_{1t,++}=\left(
\begin{array}{cc}
a_{+} & b_{+} \\
c_{+} & d_{+}%
\end{array}
\right) ,\bar{H}_{1t,--}=\bar{H}_{1t,++}^{\dag},  \label{3dH1tpp} \\
\bar{H}_{1t,+-}=\left(
\begin{array}{cc}
e_{1} & f_{-} \\
f_{+} & e_{2}%
\end{array}
\right) , \bar{H}_{1t,-+}=\left(
\begin{array}{cc}
g_{1} & h_{-} \\
h_{+} & g_{2}%
\end{array}
\right) .  \label{3dH1tpm}
\end{align}
where $%
a_{+}=a_{1}+ia_{2},b_{+}=b_{1}+ib_{2},c_{+}=c_{1}+ic_{2},d_{+}=d_{1}+id_{2},f_{\pm}=f_{1}\pm if_{2},h_{\pm}=h_{1}\pm ih_{2}
$ and $%
a_{1},a_{2},b_{1},b_{2},c_{1},c_{2},d_{1},d_{2},e_{1},e_{2},f_{1},f_{2},g_{1},g_{2},h_{1},h_{2}
$ are all real parameters. In order to separate the RS modes from the other
non-RS modes, we should choose appropriate parameters, i.e., $\bar{H}_{1t}$
can't be a zero matrix. This guarantees that the low energy physics can
exhibit the RS physics.\newline

\section{Expansion for Hamiltonian}

As stated in the main text, we can directly expand the nonlocal elements in $%
\bar{H}(\mathbf{p})$, i.e., expanding $H_{3/2},R_{13}$ as the series of $\frac{%
\mathbf{p}}{m}$. For $d=3$,
\begin{equation}  \label{3dapproxR13}
R_{13}=\left(
\begin{array}{cc}
0 & \frac{{h^{\prime}}_{13}}{2m} \\
\frac{{h^{\prime}}_{13}}{2m} & 0%
\end{array}
\right) +O(\frac{p^2}{m^2}),
\end{equation}

\begin{equation}  \label{3dapproxH3}
H_{3/2}=\sigma_{1}\otimes\mathbf{\Sigma}^{\frac{3}{2}}\cdot\mathbf{p}%
+m\sigma_{3}\otimes\sigma_{0}+O(\frac{p^2}{m^2}).
\end{equation}
For $d=2$,
\begin{equation}  \label{2dapproxR13}
R_{13}=\left(
\begin{array}{cc}
\frac{p_{+}}{2m} & 0 \\
0 & \frac{p_{-}}{2m}%
\end{array}
\right) +O(\frac{p^2}{m^2}),
\end{equation}

\begin{equation}
H_{3/2}=\left(
\begin{array}{cc}
m+\frac{p^{2}}{2m} & 0 \\
0 & -m-\frac{p^{2}}{2m}%
\end{array}%
\right) +O(\frac{p^4}{m^4}).  \label{2dapproxH3}
\end{equation}%
The Hamiltonians for $\phi _{1/2}$ and $\psi _{d+1}$ take the original form
because they contain no nonlocal terms. 
For $d=3$ or $d=2$, we have given the form of $H_{1t}$ is Sec. VI.
According to the main text, $\bar{H}_{13}=-\bar{H}_{1t}R_{t1}R_{13}$ and $%
\bar{H}_{t3}=-\bar{H}_{t1}R_{13}$. Because $\bar{H}$ is Hermitian, we can
then get the expansions of all the block matrices. Although without nonlocal
terms, the Hamiltonian matrix in CMS is still very large ($6\times6$ for $%
d=2 $ while $16\times16$ for $d=3$). We just show the approximated
Hamiltonian in CMS for $d=2$ as follows,
\begin{widetext}
\begin{equation}\label{h6by6}
\left(\begin{array}{cccccc}
m+\frac{p^2}{2m}& 0& -i\frac{a_1p_-}{\sqrt2m}& -i\frac{b_-p_-}{\sqrt2m}& -\frac{a_1p_-}{2m}& -\frac{b_-p_-}{2m}\\
0& -m-\frac{p^2}{2m}&  -i\frac{b_+p_+}{\sqrt2m}& -i\frac{a_2p_+}{\sqrt2m}& -\frac{b_+p_+}{2m}& -\frac{a_2p_+}{2m}\\
i\frac{a_1p_+}{\sqrt2m}& i\frac{b_-p_-}{\sqrt2m}& m& p_-& a_1& b_-\\
i\frac{b_+p_+}{\sqrt2m}& i\frac{a_2p_-}{\sqrt2m}& p_+& -m& b_+& a_2\\
-\frac{a_1p_+}{2m}& -\frac{b_-p_-}{2m}& a_1& b_-& m& p_-\\
-\frac{b_+p_+}{2m}& -\frac{a_2p_-}{2m}& b_+& a_2& p_+& -m
\end{array}\right).
\end{equation}
\end{widetext}

\section{Approximation: perturbation theory for the constrained system}

In this section, we  use the standard perturbation theory to solve the RS eigen
problem approximately. In 2D RS systems, by requiring $\bar{H}$ contains the
approximated eigenvectors of RS equations with the approximated eigenvalues,
we can get 3$\times $3 $\bar{H}(\mathbf{p})$ by separation of the positive
and negative energy states as shown later. No matter for 2D or 3D RS systems, we conclude that we would get three equations for
stationary states as follows,
\begin{equation}
h_{33}\phi _{3/2}+h_{31}\phi _{1/2}=\epsilon \phi _{3/2},  \label{RS3}
\end{equation}%
\begin{equation}
h_{13}\phi _{3/2}+h_{11}\phi _{1/2}=\epsilon \phi _{1/2},  \label{RS1}
\end{equation}%
\begin{equation}
H_{1/2}\phi _{1/2}=\epsilon \phi _{1/2}.  \label{RSt}
\end{equation}%
With the representation transformation $R$, we can use $\phi _{3/2},\phi _{1/2}$
and $\psi _{d+1}$ to represent the vector spinor $\psi _{\mu }$. In the rest
frame, i.e. $\mathbf{p}=(0,0,0)$ or $(0,0)$ for $d=3$ or $2$, the eigen
solutions can be easily got and are denoted as $\epsilon _{\pm }^{(0)}=\pm m$%
. The corresponding eigenvectors are $(u_{\pm,3/2 }^{(0)},u_{\pm,1/2
}^{(0)},u_{t,\pm }^{(0)})^{T}$ where $u_{\pm ,1/2}^{(0)}=u_{\pm
,t}^{(0)}=(0,0)^{T}$ or $(0,0,0,0)^{T}$. We note that $%
u_{+,3/2}^{(0)}=(1,0)^{T}$ and $u_{-,3/2}^{(0)}=(0,1)^{T}$ for 2D systems and $%
u_{+,3/2}^{(0)}=e_{1},e_{2},e_{3},e_{4}$,$%
u_{-,3/2}^{(0)}=e_{5},e_{6},e_{7},e_{8} $ for 3D systems, where $e_{i}$ is the
$i$th column of $I_{8\times 8}$. In the spirit of perturbation theory,
we write (\ref{RS3}-\ref{RSt}) to the first order of $\frac{\mathbf{p}}{m}$,
\begin{equation}
h_{33}^{(0)}u_{\pm ,3/2}^{(1)}+h_{33}^{(1)}u_{\pm ,3/2}^{(0)}=\epsilon _{\pm
}^{(0)}u_{\pm ,3/2}^{(1)}+\epsilon _{\pm }^{(1)}u_{\pm ,3/2}^{(0)},
\label{RS3a1}
\end{equation}%
\begin{equation}
h_{13}^{(1)}u_{\pm ,3/2}^{(0)}+h_{11}^{(0)}u_{\pm ,1/2}^{(1)}=\epsilon _{\pm
}^{(0)}u_{\pm ,1/2}^{(1)},  \label{RS1a1}
\end{equation}%
\begin{equation}
H_{1/2}^{(0)}u_{\pm ,1/2}^{(1)}=\epsilon _{\pm }^{(0)}u_{\pm ,1/2}^{(1)}.
\label{RSta1}
\end{equation}%
By multiplying $u_{\pm ,3/2}^{(0)}$ to the left side of (\ref{RS3a1}), we find
that $\epsilon _{\pm }^{(1)}=0$ and multiplying $u_{\mp ,3/2}^{(0)}$ to the
left of (\ref{RS3a1}), we find that $u_{\pm ,3/2}^{(1)}=\pm \frac{%
h_{33}^{(1)}u_{\pm ,3/2}^{(0)}}{2m}$.

Similarly, according to (\ref{RS1a1}) and (\ref{RSta1}), we find that, $%
u_{\pm,1/2}^{(1)}=\pm\frac{h^{(1)}_{13}u_{\pm,3/2}^{(0)}}{2m}$. To the second
order, the eigenvalues are $\epsilon^{(2)}_{\pm}=\pm\frac{p^{2}}{2m}$.\\

Finally, for the 3D RS systems, to the second order , the eigenvalues become
$m+\frac{p^{2}}{2m}$ and $-m-\frac{p^{2}}{2m}$. The corresponding
eigenvectors, to the first order in $\frac{\mathbf{p}}{m}$, are listed in
the following table where ${e^{\prime }_i}^{\prime }s$($i=1,2,3,4$) are the $%
i$th column matrix of $I_{4\times4}$:

\begin{widetext}
\centering
\begin{tabular}{|c|c|c|}
\hline
$u_{+,3/2}$ & $u_{-,3/2}$ & $u_{+,1/2}$\\
\hline
$(1,0,0,0,(\frac
{\mathbf{\Sigma}^{\frac{3}{2}}\cdot\mathbf{p}}{2m}e_1')^T)^{T}$& $(-(\frac{\mathbf{\Sigma}^{\frac{3}{2}}\cdot\mathbf{p}}%
{2m}e'_1)T,1,0,0,0)^T$& $(0,0,(\frac{{h^{\prime}}_{13}}{2m}%
e'_1)^T)^T$  \\
\hline
$(0,1,0,0,(\frac
{\mathbf{\Sigma}^{\frac{3}{2}}\cdot\mathbf{p}}{2m}e'_2)^{T})^{T}$ & $(-(\frac{\mathbf{\Sigma}^{\frac{3}{2}}\cdot\mathbf{p}}%
{2m}e_2')^{T},0,1,0,0)^T$ & $(0,0,\frac{{h^{\prime}}_{13}}{2m}%
e_2')^{T})^T$  \\
\hline
$(0,0,1,0,(\frac
{\mathbf{\Sigma}^{\frac{3}{2}}\cdot\mathbf{p}}{2m}e_3')^{T})^{T}$  & $(-(\frac{\mathbf{\Sigma}^{\frac{3}{2}}\cdot\mathbf{p}}%
{2m}e_3')^{T},0,0,1,0)^T$ & $(0,0,(\frac{{h^{\prime}}_{13}}{2m}%
e_3')^{T})^T$  \\
\hline
$(0,0,0,1,(\frac
{\mathbf{\Sigma}^{\frac{3}{2}}\cdot\mathbf{p}}{2m}e_4')^{T})^{T}$  & $(-(\frac{\mathbf{\Sigma}^{\frac{3}{2}}\cdot\mathbf{p}}%
{2m}e_4')^{T},0,0,0,1)^T$ & $(0,0,(\frac{{h^{\prime}}_{13}}{2m}%
e_4')^{T})^T$  \\
\hline
$u_{-,1/2}$ & $u_{+,t}$ & $u_{-,t}$ \\
\hline
$(-(\frac{{h^{\prime}}_{13}}{2m}e_1')^{T},0,0)^T$ & $(-i\sqrt3(\frac{{h^{\prime}%
}_{13}}{2m}e_1')^{T},0,0)^T$ & $(0,0,i(\sqrt
3\frac{{h^{\prime}}_{13}}{2m}e_1')^{T})^T$ \\
\hline
$(-(\frac{{h^{\prime}}_{13}}{2m}e_2')^{T},0,0)^T$ & $(-i\sqrt3(\frac{{h^{\prime}%
}_{13}}{2m}e_2')^{T},0,0)^T$ & $(0,0,i\sqrt
3(\frac{{h^{\prime}}_{13}}{2m}e_2')^{T})^T$ \\
$(-(\frac{{h^{\prime}}_{13}}{2m}e_3')^{T},0,0)$ & $(-i\sqrt3(\frac{{h^{\prime}%
}_{13}}{2m}e_3')^{T},0,0)^T$ & $(0,0,i\sqrt
3(\frac{{h^{\prime}}_{13}}{2m}e_3')^{T})^T$ \\
\hline
$(-(\frac{{h^{\prime}}_{13}}{2m}e_4')^{T},0,0)^T$ & $(-i\sqrt3(\frac{{h^{\prime}%
}_{13}}{2m}e_4')^{T},0,0)^T$ & $(0,0,i\sqrt
3(\frac{{h^{\prime}}_{13}}{2m}e_4')^{T})^T$ \\
\hline
\end{tabular}\\
\end{widetext}

For the 2D RS systems, to the second order of $\frac{\mathbf{p}}{m} $, the
eigenvalues become $m+\frac{p^{2}}{m}$ and $-m-\frac{p^{2}}{m}$ while to the
first order of $\frac{\mathbf{p}}{m}$, the eigenvectors become, $%
u_{+,3/2}\approx(1,0),u_{-,3/2}\approx(0,1)$, $u_{+,1/2}\approx(\frac{p_{+}}{2m}%
,0),u_{-,1/2}\approx(0,\frac{p_{-}}{2m})$, $u_{+,t}\approx-i\sqrt 2(\frac{p_{+}%
}{2m},0)$ and $u_{-,t}\approx-i\sqrt2(0,\frac{p_{-}}{2m})$, respectively. We
find that under such approximations, for a finite $\mathbf{p}$, the positive
and negative energy sectors in the 2D RS systems are separated while in the
3D RS systems, they are not.

Let $H^{\prime }_{3/2}=H^{\prime }_{1/2}=H^{\prime }_{t}=$diag$(m+\frac{p^{2}}{2m%
},-m-\frac{p^{2}}{2m})$, and $R^{\prime }_{13}=$diag$(\frac{p_{+}}{2m},\frac{%
p_{-}}{2m}),R^{\prime }_{t1}=-i\sqrt2$. We can find that the eigenvectors of
$H^{\prime }_{3/2}$ is just $(1,0)^{T}$ or $(0,1)^{T}$ with eigenvalues $m+%
\frac{p^{2}}{2m}$ or $-m-\frac{p^{2}}{2m}$, respectively. Then $R^{\prime
}_{13}(1,0)^{T}$($R^{\prime }_{t1}R_{13}(1,0)^{T}$) and $R^{\prime
}_{13}(0,1)^{T}$($R^{\prime }_{t1}R_{13}(0,1)^{T}$) are the eigenvectors of $%
H^{\prime }_{1}$($H^{\prime }_{t}$) with eigenvalues $m+\frac{p^{2}}{2m}$ or
$-m-\frac{p^{2}}{2m}$, respectively. According to this fact, we then can
construct a Hamiltonian in CMS which contains the approximative solutions in
the 2D RS systems while it is difficult for three dimensions. We find that $%
\bar{H^{\prime }}_{1t}=\left(
\begin{array}{cc}
a_{1} & b_{-} \\
b_{+} & a_{2}%
\end{array}
\right) $ where $b_{\pm}=b_{1}\pm ib_{2}$ and $a_{1},a_{2},b_{1},b_{2}$ are
all real parameters. We take them real constants for simplicity. With this
choice, we have $\bar {H^{\prime }}_{31}=-i\sqrt2\left(
\begin{array}{cc}
a_{1}\frac{p_{-}}{2m} & b_{-}\frac{p_{-}}{2m} \\
b_{+}\frac{p_{+}}{2m} & a_{2}\frac{p_{+}}{2m}%
\end{array}
\right) $, and $\bar{H^{\prime }}_{3t}=\left(
\begin{array}{cc}
-a_{1}\frac{p_{-}}{2m} & -b_{-}\frac{p_{-}}{2m} \\
-b_{+}\frac{p_{+}}{2m} & -a_{2}\frac{p_{+}}{2m}%
\end{array}
\right) $. Finally, The constructed $\mathbf{k}\cdot\mathbf{p}$ Hamiltonian
in CMS is given by
\begin{widetext}
\begin{equation}\label{2dh}
\left(\begin{array}{cccccc}
m+\frac{p^2}{2m}& 0& -i\frac{a_1p_-}{\sqrt2m}& -i\frac{b_-p_-}{\sqrt2m}& -\frac{a_1p_-}{2m}& -\frac{b_-p_-}{2m}\\
0& -m-\frac{p^2}{2m}& -i\frac{b_+p_+}{\sqrt2m}& -i\frac{a_2p_+}{\sqrt2m}& -\frac{b_+p_+}{2m}& -\frac{a_2p_+}{2m}\\
i\frac{a_1p_+}{\sqrt2m}& i\frac{b_-p_-}{\sqrt2m}& m+\frac{p^2}{2m}& 0& a_1& b_-\\
i\frac{b_+p_+}{\sqrt2m}& i\frac{a_2p_-}{\sqrt2m}& 0& -m-\frac{p^2}{2m}&  b_+& a_2\\
-\frac{a_1p_+}{2m}& -\frac{b_-p_-}{2m}& a_1& b_-& m+\frac{p^2}{2m}& 0\\
-\frac{b_+p_+}{2m}&  -\frac{a_2p_-}{2m}& b_+& a_2& 0& -m-\frac{p^2}{2m}
\end{array}\right)
\end{equation}.
\end{widetext}
When $b_{1}=b_{2}=0$, we find that the positive and negative sectors are
separated. By changing the order of the basis set, the above Hamiltonian is
in a block form, $\left(
\begin{array}{cc}
H_{+} & 0 \\
0 & H_{-}%
\end{array}
\right) $, where, $H_{+}=(m+\frac{p^{2}}{2m})I_{3\times3}+a_{1}\left(
\begin{array}{ccc}
0 & -i\frac{p_{-}}{\sqrt2m} & -\frac{p_{-}}{2m} \\
i\frac{p_{+}}{\sqrt2m} & 0 & 1 \\
-\frac{p_{+}}{2m} & 1 & 0%
\end{array}
\right) $ and $H_{-}=(-m-\frac{p^{2}}{2m})I_{3\times3}+a_{2}\left(
\begin{array}{ccc}
0 & -i\frac{p_{+}}{\sqrt2m} & -\frac{p_{+}}{2m} \\
i\frac{p_{-}}{\sqrt2m} & 0 & 1 \\
-\frac{p_{-}}{2m} & 1 & 0%
\end{array}
\right) $. With respect to a unitary transformation $U$ which doesn't depend
on $\mathbf{p}$,
\begin{widetext}\begin{equation}\label{hp}
H'_+=U^\dag H_+U=(m+\frac{p^2}{2m})I_{3\times3}+a_1\left(\begin{array}{ccc}
0& -\sqrt{\frac{3}{8}}\frac{p_-}{m}&  -\sqrt{\frac{3}{8}}\frac{p_-}{m}\\
-\sqrt{\frac{3}{8}}\frac{p_+}{m}& 1& 0\\
-\sqrt{\frac{3}{8}}\frac{p_+}{m}& 0& -1
\end{array}\right),
\end{equation}\end{widetext}
and
\begin{widetext}\begin{equation}\label{hm}
H'_-=U^\dag H_-U=(-m-\frac{p^2}{2m})I_{3\times3}+a_2\left(\begin{array}{ccc}
0& -\sqrt{\frac{3}{8}}\frac{p_+}{m}&  -\sqrt{\frac{3}{8}}\frac{p_+}{m}\\
-\sqrt{\frac{3}{8}}\frac{p_-}{m}& 1& 0\\
-\sqrt{\frac{3}{8}}\frac{p_-}{m}& 0& -1
\end{array}\right),
\end{equation}
\end{widetext}
with $U=\left(
\begin{array}{ccc}
1 & 0 & 0 \\
0 & \frac{e^{-i\theta}}{\sqrt2} & -\frac{e^{i\theta}}{\sqrt2} \\
0 & \frac{e^{-i\theta}}{\sqrt2} & \frac{e^{i\theta}}{\sqrt2}%
\end{array}
\right) $ and $\arctan\theta=\sqrt2$.

Notice that we have used the units $c=\hbar=1$. After restoring the units,
we have
\begin{widetext}\begin{equation}\label{hpm}
H'_{\pm}=(\pm\frac{\hbar^2q^2}{2m})I_{3\times3}+a_{\pm}\left(\begin{array}{ccc}
0& -\sqrt{\frac{3}{8}}\frac{\hbar q_\mp}{mc}& -\sqrt{\frac{3}{8}}\frac{\hbar q_\mp}{mc}\\
-\sqrt{\frac{3}{8}}\frac{\hbar q_\pm}{mc}& 1& 0\\
-\sqrt{\frac{3}{8}}\frac{\hbar q_\pm}{mc}& 0& -1
\end{array}\right),
\end{equation}\end{widetext}
where we have neglected the energy constant $\pm m$, $a_+=a_1, a_-=a_2$ and $%
\mathbf{p}=\hbar\mathbf{q}$. 
Denoting $\Delta_{1}=\pm\frac{\hbar^{2}}{2m},\Delta_{2}=-a_{\pm}\sqrt{\frac{3%
}{8}}\frac{\hbar}{mc},\Delta_{3}=a_{\pm}$, the above Hamiltonians are
rewritten as
\begin{widetext}\begin{equation}\label{hpma}
H'_{\pm}=\left(\begin{array}{ccc}
\Delta_1(q_x^2+q_y^2)& \Delta_2q_{\mp}&  \Delta_2q_{\mp}\\
\Delta_2q_{\pm}& \Delta_1(q_x^2+q_y^2)+\Delta_3& 0\\
\Delta_2q_{\pm}& 0& \Delta_1(q_x^2+q_y^2)-\Delta_3
\end{array}\right).
\end{equation}\end{widetext}
Interestingly, in terms of $\Delta_{i}$, we can get the effective mass and
light velocity that, $m=|\frac{\hbar^{2}}{2\Delta_{1}}|$ and $c=|\sqrt\frac{3%
}{2}\frac{\Delta_{1}\Delta_{3}}{\Delta_{2}\hbar}|$.

\section{Berry curvature and vanishing orbital magnetic moment}

In this section, we first give the details of calculating the Berry
curvature and orbital magnetic moment of the RS band in $H_{+}^{\prime }(%
\mathbf{q})\ $of Eq. (\ref{hpma}) (note that the procedure for $H_{-}^{\prime
}(q)$ is similar).\newline

Diagonalizing $H_{+}^{\prime }$, we obtain its eigenvalues and eigenvectors
as follows (assume $\Delta _{3}>0$. Once $\Delta _{3}<0$, we can exchange
the second and third basis vectors to make $\Delta _{3}>0$): The eigenvalues
are $\epsilon _{RS}=\Delta _{1}q^{2}$ (RS band) and $\epsilon _{\pm }=\Delta
_{1}q^{2}\pm \sqrt{\Delta _{3}^{2}+2\Delta _{2}^{2}q^{2}}$ (non-RS bands),
while the corresponding eigenvectors are then, $u_{RS}=N_{RS}(1,-\frac{%
\Delta _{2}}{\Delta _{3}}q_{+},\frac{\Delta _{2}}{\Delta _{3}}q_{+})^{T}$, $%
u_{+}=N_{+}(\frac{2\Delta _{2}q_{-}}{\Delta _{3}+\lambda (q)},1,\frac{%
\lambda (q)-\Delta _{3}}{\lambda (q)+\Delta _{3}})^{T}$ and $u_{-}=N_{-}(%
\frac{2\Delta _{2}q_{-}}{\Delta _{3}+\lambda (q)},-\frac{\lambda (q)-\Delta
_{3}}{\lambda (q)+\Delta _{3}},-1)^{T}$ where $\lambda (q)=\sqrt{\Delta
_{3}^{2}+2\Delta _{2}^{2}q^{2}}$ and $N_{RS},N_{\pm }$ are the normalization
factors, $N_{RS}=\frac{1}{\sqrt{1+2\frac{\Delta _{2}^{2}q^{2}}{\Delta
_{3}^{2}}}},N_{\pm }=\frac{\lambda +\Delta _{3}}{2\lambda }$. \newline

Then we can make a direct calculation according to $\Omega
_{n}^{z}=i(<\partial _{q_{x}}u_{n}|\partial _{q_{y}}u_{n}>-c.c.)$ and $%
m_{n}^{z}=-\frac{ie}{2}(<\partial _{q_{x}}u_{n}|H-\epsilon _{n}|\partial
_{q_{y}}u_{n}>-c.c.)$\cite{rmp} where $\Omega _{n}^{z}$ and $m_{n}^{z}$ are
the Berry curvature and orbital magnetic moment for the $n$th band
respectively while $H(\mathbf{q})$ and $\epsilon _{n}(\mathbf{q})$ are the
Bloch Hamiltonian and the eigen-energies for the $n$th band, respectively.
We then find that
\begin{equation*}
\Omega _{RS}^{z}=-\frac{4\Delta _{2}^{2}\Delta _{3}^{2}}{(\Delta
_{3}^{2}+2\Delta _{2}^{2}q^{2})^{2}}=-2\Omega _{+}^{z}=-2\Omega _{-}^{z},
\end{equation*}%
\begin{equation*}
m_{RS}^{z}=0,m_{+}^{z}=\frac{e\Delta _{2}^{2}\Delta _{3}^{2}}{(\Delta
_{3}^{2}+2\Delta _{2}^{2}q^{2})^{3/2}}=-m_{-}^{z},
\end{equation*}%
where the orbital magnetic moment for the RS band is always vanishing while
its Berry curvature is not. This is totally different from the case for the
massive Dirac quasiparticles where the orbital magnetic moment is
proportional to the Berry curvature \cite{rmp}. Actually the peculiarity is
originated from the non-trivial constraints in the RS equations. Another
form of the formulas for $\Omega _{n}^{z}$ and $m_{n}^{z}$ will help us to
understand this peculiarity:
\begin{equation}
\Omega _{n}^{z}=i\sum\limits_{n^{\prime }}[\frac{<u_{n}|\partial
_{q_{x}}H|u_{n^{\prime }}><u_{n^{\prime }}|\partial _{q_{y}}H|u_{n}>}{%
(\epsilon _{n}-\epsilon _{n^{\prime }})^{2}}-c.c.],  \label{omega}
\end{equation}%
\begin{equation}
m_{n}^{z}=\frac{ie}{2}\sum\limits_{n^{\prime }}(\frac{<u_{n}|\partial
_{q_{x}}H|u_{n^{\prime }}><u_{n^{\prime }}|\partial _{q_{y}}H|u_{n}>}{%
\epsilon _{n}-\epsilon _{n^{\prime }}}-c.c.).  \label{m}
\end{equation}%
As $\partial _{q_{x}}H_{+}^{\prime }=2\Delta _{1}q_{x}I_{3\times 3}+\Delta
_{2}\left(
\begin{array}{ccc}
0 & 1 & 1 \\
1 & 0 & 0 \\
1 & 0 & 0%
\end{array}%
\right) $, $\partial _{q_{y}}H_{+}^{\prime }=2\Delta _{1}q_{y}I_{3\times
3}+\Delta _{2}\left(
\begin{array}{ccc}
0 & -i & -i \\
i & 0 & 0 \\
i & 0 & 0%
\end{array}%
\right) $, we will get,
\begin{equation}
<u_{RS}|\partial _{q_{x}}H_{+}^{\prime }|u_{+}>=\frac{\Delta _{2}}{\sqrt{1+%
\frac{2\Delta _{2}^{2}q^{2}}{\Delta _{3}^{2}}}},  \label{xRSp}
\end{equation}%
\begin{equation}
<u_{RS}|\partial _{q_{y}}H_{+}^{\prime }|u_{+}>=\frac{-i\Delta _{2}}{\sqrt{1+%
\frac{2\Delta _{2}^{2}q^{2}}{\Delta _{3}^{2}}}},  \label{yRSp}
\end{equation}%
\begin{equation}
<u_{RS}|\partial _{q_{x}}H_{+}^{\prime }|u_{-}>=-\frac{\Delta _{2}}{\sqrt{1+%
\frac{2\Delta _{2}^{2}q^{2}}{\Delta _{3}^{2}}}},  \label{xRSm}
\end{equation}%
\begin{equation}
<u_{RS}|\partial _{q_{y}}H_{+}^{\prime }|u_{-}>=\frac{i\delta _{2}}{\sqrt{1+%
\frac{2\Delta _{2}^{2}q^{2}}{\Delta _{3}^{2}}}}.  \label{yRSm}
\end{equation}%
Then it is easy to find that,
\begin{equation}
<u_{RS}|\partial _{q_{x}}H_{+}^{\prime }|u_{+}><u_{+}|\partial
_{q_{y}}H_{+}^{\prime }|u_{RS}>=<u_{RS}|\partial _{q_{x}}H_{+}^{\prime
}|u_{-}><u_{-}|\partial _{q_{y}}H_{+}^{\prime }|u_{RS}>=\frac{i\Delta
_{2}^{2}}{1+2\frac{\Delta _{2}^{2}q^{2}}{\Delta _{3}^{2}}}.  \label{product}
\end{equation}%
According to Eq. (\ref{omega}), we can write $\Omega _{RS}^{z}$ as the sum
of the contributions from $u_{+}$ and $u_{-}$:
\begin{equation*}
\Omega _{RS}^{z}=\Omega _{RS,+}^{z}+\Omega _{RS,-}^{z}
\end{equation*}%
where, $\Omega _{RS,+}^{z}=-2Im[\frac{<u_{RS}|\partial _{q_{x}}H_{+}^{\prime
}|u_{+}><u_{+}|\partial _{q_{y}}H_{+}^{\prime }|u_{RS}>}{(\epsilon
_{RS}-\epsilon _{+})^{2}}]$, and $\Omega _{RS,-}^{z}=-2Im[\frac{%
<u_{RS}|\partial _{q_{x}}H_{+}^{\prime }|u_{-}><u_{-}|\partial
_{q_{y}}H_{+}^{\prime }|u_{RS}>}{(\epsilon _{RS}-\epsilon _{-})^{2}}]$. They
are equal, because according to Eq. (\ref{product}), the denominators of $%
\Omega _{RS,+}^{z}$ and $\Omega _{RS,-}^{z}$ are the same while the
numerators, i.e., $(\epsilon _{RS}-\epsilon _{\pm })^{2}$, are also the
same. The contributions from $u_{\pm }$ are additive, i.e., $\Omega
_{RS,+}^{z}=\Omega _{RS,-}^{z}=-\frac{2\Delta _{2}^{2}\Delta _{3}^{2}}{%
(\Delta _{3}^{2}+2\Delta _{2}^{2}q^{2})^{2}}$ , then, $\Omega _{RS}^{z}=-%
\frac{4\Delta _{2}^{2}\Delta _{3}^{2}}{(\Delta _{3}^{2}+2\Delta
_{2}^{2}q^{2})^{2}}$.

However, the contributions from $u_{\pm }$ to the orbital magnetic moment of
the RS band are subtractive. According to Eq. (\ref{m}),
\begin{equation*}
m_{RS}^{z}=m_{RS,+}^{z}+m_{RS,-}^{z}
\end{equation*}%
where, $m_{RS,+}^{z}=-eIm(\frac{<u_{RS}|\partial _{q_{x}}H_{+}^{\prime
}|u_{+}><u_{+}|\partial _{q_{y}}H_{+}^{\prime }|u_{RS}>}{\epsilon
_{RS}-\epsilon _{+}})$, and $m_{RS,-}^{z}=-eIm(\frac{<u_{RS}|\partial
_{q_{x}}H_{+}^{\prime }|u_{-}><u_{-}|\partial _{q_{y}}H_{+}^{\prime }|u_{RS}>%
}{\epsilon _{RS}-\epsilon _{-}})$. They are opposite in sign, because
according to Eq. (\ref{product}), the denominators of $m_{RS,+}^{z}$ and $%
m_{RS,-}^{z}$ are the same while the numerators, i.e., $\epsilon
_{RS}-\epsilon _{\pm }$, are opposite in sign. Then $m_{RS,+}^{z}=-\frac{%
2e\Delta _{2}^{2}\Delta _{3}^{2}}{(\Delta _{3}^{2}+2\Delta
_{2}^{2}q^{2})^{3/2}}=-m_{RS,-}^{z}$, so $m_{RS}^{z}=0$.\newline

We now try to give the generic criterion for the vanishing orbital magnetic
moment. One can always rewrite an arbitrary $s\times s$ Hamiltonian $%
H_{s\times s}(\mathbf{q})$ ($s=2N+1$, $N>0$ is an integer) as $H_{s\times s}(%
\mathbf{q})=h_{s\times s}(\mathbf{q})+C(\mathbf{q})I_{s\times s}$ ($C(%
\mathbf{q})$ is just a scalar quantity). Assuming that we can find a
constant unitary matrix $T_{s\times s}$ so that $h(\mathbf{q})$ satisfies

\begin{equation}  \label{condition}
T^{\dag}h(\mathbf{q})T=-h(\mathbf{q}),
\end{equation}

and there is no degeneracy. Eq. (\ref{condition}) means that

\begin{equation}  \label{condition1}
T^{\dag}(H(\mathbf{q})-C(\mathbf{q}))T=-(H(\mathbf{q})-C(\mathbf{q})),
\end{equation}
i.e.,

\begin{equation}  \label{condition2}
T^{\dag}H(\mathbf{q})T=2C(\mathbf{q})-H(\mathbf{q}).
\end{equation}

Suppose $|u>$ is the eigenvector of $H$, i.e., $H|u>=\epsilon |u>$, then $%
T|u>$ would also be an eigenvector, $HT|u>=(2C-\epsilon )T|u>$ according to
Eq. (\ref{condition2}). Then the eigenvectors will be coming in pairs while
the one with eigenvalue $C$ has no partner.%
Hence, the eigenvectors can be labeled by $N_{3}$: $H|u_{N_{3}}>=\epsilon
_{N_{3}}|u_{N_{3}}>$ where $N_{3}=-N,-N+1,\ldots ,N-1,N$ with $\epsilon
_{-N_{3}}=2C-\epsilon _{N_{3}}$ and T connects a state $|u>$ with its
partner, namely,

\begin{equation}
T|u_{N_{3}}>=|u_{-N_{3}}>.  \label{condition5}
\end{equation}%
Consequently,
\begin{equation}
T^{\dag }(H-\epsilon _{N_{3}})T=-(H-\epsilon _{-N_{3}}).
\end{equation}%
Thus the orbital magnetic moments are restricted to satisfy $%
m_{-N_{3}}^{z}=-m_{N_{3}}^{z}$.
The reason is that, 
\begin{equation*}
m_{-N_{3}}^{z}=-\frac{ie}{2}(<\partial _{q_{x}}u_{-N_{3}}|H(\mathbf{q}%
)-\epsilon _{-N_{3}}(\mathbf{q})|\partial _{q_{y}}u_{-N_{3}}>-c.c.)=
\end{equation*}%
\begin{equation*}
\frac{ie}{2}(<\partial _{q_{x}}u_{-N_{3}}|T^{\dag }(H(\mathbf{q})-\epsilon
_{N_{3}}(\mathbf{q}))T|\partial _{q_{y}}u_{-N_{3}}>-c.c.)=
\end{equation*}%
\begin{equation*}
\frac{ie}{2}(<\partial _{q_{x}}u_{N_{3}}|H(\mathbf{q})-\epsilon _{N_{3}}(%
\mathbf{q})|\partial _{q_{y}}u_{N_{3}}>-c.c.)=-m_{N_{3}}^{z},
\end{equation*}%
where we have used the relation $T\partial _{\mathbf{q}}|u_{N_{3}}>=\partial
_{\mathbf{q}}|u_{-N_{3}}>$ according to Eq. (\ref{condition5}).

Thus $m^z_0=0$, i.e., the orbital magnetic moment of the middle band is
restricted to be vanishing. On the other hand, it is similar to find that $%
\Omega^z_{-N_3}=\Omega^z_{N_3}$, which won't restrict $\Omega_0^z$ to be
zero.\newline

We can also write $m_{0}^{z}$ as the contributions from the other bands,
i.e.,
\begin{equation*}
m_{0}^{z}=(m_{0,+1}^{z}+m_{0,-1}^{z})+(m_{0,+2}^{z}+m_{0,-2}^{z})+\ldots
+(m_{0,+N}^{z}+m_{0,-N}^{z}).
\end{equation*}%
For each parenthesis $(m_{0,N^{\prime }}^{z},m_{0,-N^{\prime }}^{z})$ where $%
N^{\prime }=1,2,\ldots ,N$, $m_{0,N^{\prime }}^{z}=-m_{0,-N^{\prime }}^{z}$,
then they are subtractive and $m_{0}^{z}=0$. This is because,
\begin{equation*}
m_{0,N^{\prime }}^{z}=-eIm(\frac{<u_{0}|\partial _{q_{x}}H|u_{N^{\prime
}}><u_{N^{\prime }}|\partial _{q_{y}}H|u_{0}>}{\epsilon _{0}-\epsilon
_{N^{\prime }}}),
\end{equation*}%
\begin{equation*}
m_{0,-N^{\prime }}^{z}=-eIm(\frac{<u_{0}|\partial _{q_{x}}H|u_{-N^{\prime
}}><u_{-N^{\prime }}|\partial _{q_{y}}H|u_{0}>}{\epsilon _{0}-\epsilon
_{-N^{\prime }}}).
\end{equation*}%
As $\epsilon _{0}=C,\epsilon _{N^{\prime }}=2C-\epsilon _{-N^{\prime }}$,
the numerators, i.e., $\epsilon _{0}-\epsilon _{-N^{\prime }}=-(\epsilon
_{0}-\epsilon _{N^{\prime }})$. As $u_{-N^{\prime }}=Tu_{N^{\prime
}},u_{0}=Tu_{0}$, then, the denominators,
\begin{equation*}
<u_{0}|\partial _{q_{x}}H|u_{-N^{\prime }}><u_{-N^{\prime }}|\partial
_{q_{y}}H|u_{0}>=<u_{0}|T^{\dag }\partial _{q_{x}}HT|u_{N^{\prime
}}><u_{N^{\prime }}|T^{\dag }\partial _{q_{y}}HT|u_{0}>.
\end{equation*}%
According to Eq. (\ref{condition2}), $T^{\dag }\partial _{q_{x}}HT=2\partial
_{q_{x}}C-\partial _{q_{x}}H$,$T^{\dag }\partial _{q_{y}}HT=2\partial
_{q_{y}}C-\partial _{q_{y}}H$, then,
\begin{equation*}
<u_{0}|\partial _{q_{x}}H|u_{-N^{\prime }}><u_{-N^{\prime }}|\partial
_{q_{y}}H|u_{0}>=<u_{0}|\partial _{q_{x}}H|u_{N^{\prime }}><u_{N^{\prime
}}|\partial _{q_{y}}H|u_{0}>,
\end{equation*}%
where we have used the relation that $<u_{0}|u_{N^{\prime }}>=0$. Finally, $%
m_{0,N^{\prime }}^{z}=-m_{0,-N^{\prime }}^{z}$.\newline

For $H_{+}^{\prime }=\Delta _{1}q^{2}I_{3\times 3}+h$ in Eq. (\ref{hpma}),
we can find a $T$ matrix, $T=\left(
\begin{array}{ccc}
1 & 0 & 0 \\
0 & 0 & -1 \\
0 & -1 & 0%
\end{array}%
\right) $ so that $T^{\dag }hT=-h$, and $C(\mathbf{q})=\Delta _{1}q^{2}$. We
can then label the eigenvalues and eigenvectors of $H_{+}^{\prime }$ by $%
N_{3}=0,\pm $, i.e., $u_{RS}\equiv u_{0}$ while $\epsilon _{RS}\equiv
\epsilon _{0}$. It is easy to find that $Tu_{\pm }=u_{\mp }$ and $%
Tu_{RS}=u_{RS}$. According to the above arguments, we can find that $%
m_{RS}^{z}$ is restricted to be zero while $\Omega _{RS}^{z}$ is not, which
is consistent with the direct calculation.

\section{material realizations: symmetry requirements}

In order to get the 2D RS excitations in CMS, we firstly consider the
generic Hamiltonian (Eq. (\ref{hpma})) in the following form:
\begin{widetext}\begin{equation}\label{rsh}
\mathcal{H}(q_x,q_y)=\left(\begin{array}{ccc}
\Delta_1(q_x^2+q_y^2)+E_1& \Delta_2q_{\mp}& \Delta_2q_{\mp}\\
\Delta_2q_{\pm}&  \Delta_1(q_x^2+q_y^2)+E_2&   \Delta_{23}\\
\Delta_2q_{\pm}& \Delta_{23}& \Delta_1(q_x^2+q_y^2)+E_3
\end{array}\right),
\end{equation}
\end{widetext}
where $q_{\pm}=q_{x}\pm iq_{y}$ and $\Delta_{1},\Delta_{2},\Delta
_{23},E_{1},E_{2},E_{3}$ are all real parameters with appropriate units. As
we know, symmetries of Hamiltonian always add some restrictions to the form
of the low energy $\mathbf{k\cdot p}$ Hamiltonian around the point of
symmetry $\mathbf{k}^{*}$ in the Brilloin zone (BZ), because the Bloch
Hamiltonian $H(\mathbf{k})$ satisfies \cite{kp},
\begin{equation}  \label{HkRes}
D(\{\alpha|\mathbf{t}\})^{\dag}H(\alpha\mathbf{k})D(\{\alpha |\mathbf{t}%
\})=H(\mathbf{k}),
\end{equation}
where $\{\alpha|\mathbf{t}\}$ is an element of the space group, and $%
D(\{\alpha|\mathbf{t\})}$ is a matrix representation. It is obvious that any
operation in the translation group, namely $\{E|\mathbf{R}_{l}\}$ in $T$,
will not produce any restriction according to Eq. ($\ref{HkRes}$). For $%
\mathbf{k}^{*} $, any other symmetry operations $\{\beta|\mathbf{t}%
^{\prime}\}$ will produce symmetry restrictions to $H(\mathbf{k}^{*})$ when $%
\beta\mathbf{k}^{*}=\mathbf{k}^{*}+\mathbf{G}$ where $\mathbf{G}$ is a
reciprocal lattice vector. The rest symmetry operations in the space group
will generate a wave vector star, and relate the low energy models around
these vectors in the star with each other. Writing $\mathbf{q}=\mathbf{k}-%
\mathbf{k}^{*}$, and for $\{\beta|\mathbf{t}^{\prime}\}$, we can find that,
\begin{equation}  \label{HkpRes}
D(\{\beta|\mathbf{t}^{\prime}\})^\dag H(\mathbf{k}^{*}+\mathbf{G}+\beta%
\mathbf{q})D(\{\beta|\mathbf{t}^{\prime}\})=H(\mathbf{k}^{*}+\mathbf{q}).
\end{equation}
As $H(\mathbf{k}+\mathbf{G})$ is related to $H(\mathbf{k})$ by a unitary
transformation which can be absorbed in $D$, we can get the following
equation,
\begin{equation}  \label{HqRes}
\mathcal{D}(\{\beta|\mathbf{t}^{\prime}\})^\dag\mathcal{H}(\beta\mathbf{q})%
\mathcal{D}(\{\beta|\mathbf{t}^{\prime}\})=\mathcal{H}(\mathbf{q}),
\end{equation}
where $\mathcal{H}(\mathbf{q})\equiv H(\mathbf{k}^{*}+\mathbf{q})$ and $%
\mathcal{D}$ is the representation of the symmetry group of $\mathbf{k}^{*}$%
, which can be derived based on group theory.\newline

Based on the above equation, we will reveal the symmetry requirements for
getting the low energy Hamiltonian $\mathcal{H}(\mathbf{q})$ as shown in Eq.
(\ref{rsh}) step by step. Inspired by the 2D RS equations which describes
that three Dirac spinors obey Dirac equation with nontrivial constraints,
for the material realizations, we can consider three 2D layers each of which
will provide a $q^{2}$-like excitation firstly, and then consider the
coupling between these layers. We consider nonmagnetic materials with
negligible spin orbit coupling for simplicity.\newline

Around $z$-axis, $\mathbf{k}^{\ast }$ of a layer can be subject to $C_{n}$
point symmetry cyclic group \cite{group}, whose irreducible representation
(IR) can be denoted by $l_{z}$ where $l_{z}=1,2,\ldots ,n$ and the
corresponding transformation matrix is just a number: $\mathcal{D}%
(c_{n})=e^{-i\frac{2\pi l_{z}}{n}}$ where $c_{n}$ denotes the rotation
around $z$-axis by an angle $\frac{2\pi}{n}$. Using Eq. (\ref{HqRes}), we
can find that $n=3,4,6$, because $C_{1}$ would allow the first order of $%
\mathbf{q}$ while $C_{2}$ would allow the different coefficients before $%
q_{x}^{2}$ and $q_{y}^{2}$. We choose the basis sets located at the three
layer respectively, i.e., the one in the middle or second layer corresponds
to the first basis vector while the one in the first (third) layer
corresponds to the second (third) basis vector. The relatively large
distance between the first and third layers guarantees that the coupling
between the corresponding basis vectors is small, i.e., $\Delta _{23}$ can
be negligible which is required by the RS physics.\newline

\begin{widetext}\begin{figure}
\centering
\includegraphics[width=18cm]{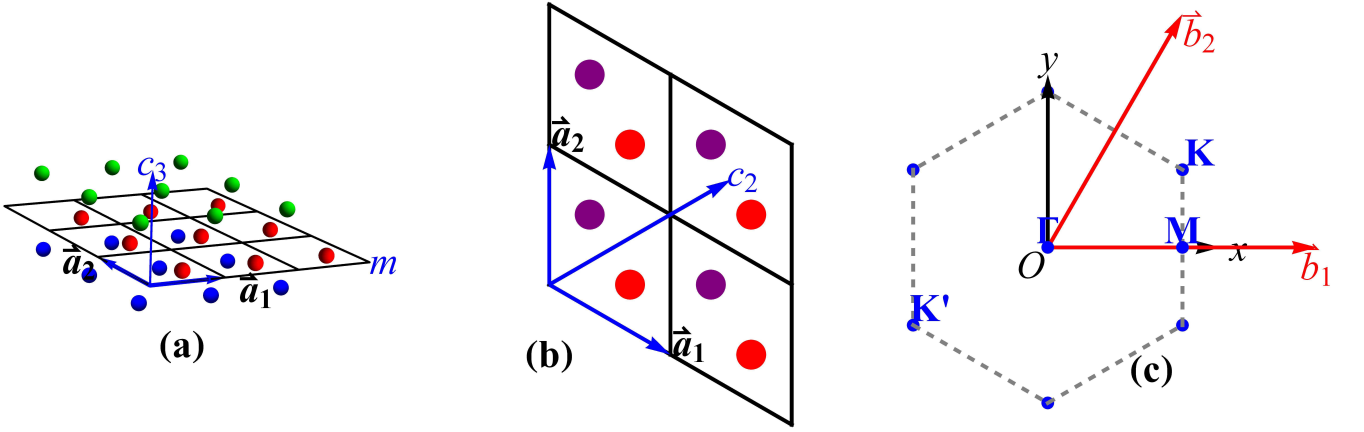}\\
\caption{In (a), we show the prototypical structure with a triangular lattice. ${\bf a}_1$ and ${\bf a}_2$ are primitive lattice basis vectors, and $c_3$ is along the $z-$ axis which is perpendicular to the both ${\bf a}_1$ and ${\bf a}_2$. The first layer is depicted in green, the second layer is depicted in red, while the third layer is depicted in  blue, and each ball represents an atom. (b) shows the top view, where the red disks shows the atoms in the second layer while the purple ones are the atoms in the first or the third layers. The second layer is a mirror plane which relate the first and third layer. In (c), we show the hexagonal BZ where ${\bf b_1}$ and ${\bf b}_2$ are reciprocal basis vectors. In (b), we also show a rotation $c_2$ which is not the symmetry operation of the lattice in (a), but can relate the first and second layer when located at the middle plane between the first and second layer.}\label{sm}
\end{figure}
\end{widetext}

Next we consider the simulation of $\mathcal{H}_{12}$ and $\mathcal{H}_{13}$%
. According to Eq. (\ref{HqRes}), we can find that, for the coupling $%
\mathcal{H}_{12}=\mathcal{H}_{13}\sim q_{\mp }$, the IR in the second layer $%
l_{z}^{2}$ and in the first (third) layer $l_{z}^{1}$ ($l_{z}^{3}$) should
satisfy $l_{z}^{2}-l_{z}^{1}=l_{z}^{2}-l_{z}^{3}=\pm 1$. The realization of
such a relation between the IR's in different layers is subtle because it is
through the displacement in the $xy$-plane between the different layers,
which guarantees that the IR's are different. We denote such a displacement
from the second layer to the first (third) layer as $\boldsymbol{\delta }$ ($%
\boldsymbol{\delta }^{\prime }$), it is easy to show that for $c_{n}$(with
respect to the second layer, i.e., the fixed point is at some atom in the
second layer) to survive, $c_{n}\boldsymbol{\delta }-\boldsymbol{\delta }=%
\mathbf{R}_l,c_{n}\boldsymbol{\delta }^{\prime }-\boldsymbol{\delta }%
^{\prime }=\mathbf{R}_{l}^{\prime }$ where $\mathbf{R}_{l},\mathbf{R}%
_{l}^{\prime }$ are lattice vectors. 
Suppose the Bloch function $\psi _{\mathbf{k}^{\ast }}(\mathbf{r})$ (which
satisfies $\psi_{\mathbf{k}^*}(\mathbf{r}-\mathbf{R}_l)=e^{-i\mathbf{k}%
^*\cdot\mathbf{R}_l}\psi_{\mathbf{k}^*}(\mathbf{r})$) is the basis vector in
the second layer with the property that $c_{n}\psi _{\mathbf{k}^{\ast }}(%
\mathbf{r})=\psi _{\mathbf{k}^{\ast }}(c_{n}^{-1}\mathbf{r})=\chi \psi _{%
\mathbf{k}^{\ast }}(\mathbf{r})$. Then in the first (third) layer, we will
have $\psi _{\mathbf{k}^{\ast }}(\mathbf{r}-\boldsymbol{\delta })$ ( $\psi _{%
\mathbf{k}^{\ast }}(\mathbf{r}-\boldsymbol{\delta ^{\prime }})$), which
satisfies that,
\begin{equation}
c_{n}\psi _{\mathbf{k}^{\ast }}(\mathbf{r}-\boldsymbol{\delta }(\boldsymbol{%
\delta }^{\prime }))=\psi _{\mathbf{k}^{\ast }}(c_{n}^{-1}\mathbf{r}-%
\boldsymbol{\delta }(\boldsymbol{\delta }^{\prime}))= e^{i\mathbf{G}\cdot
\boldsymbol{\delta }(\boldsymbol{\delta }^{\prime })}\chi \psi _{\mathbf{k}%
^{\ast }}(\mathbf{r}-\boldsymbol{\delta }(\boldsymbol{\delta }^{\prime })),
\label{undercn}
\end{equation}%
where $c_{n}\mathbf{k}^{\ast }=\mathbf{k}^{\ast }+\mathbf{G}$. Therefore, we
will have,

\begin{equation}  \label{delta}
\mathbf{G}\cdot\boldsymbol{\delta}=\mathbf{G}\cdot\boldsymbol{\delta }%
^{\prime}=\pm\frac{2\pi}{n},
\end{equation}
which rules out the $\Gamma$ point and require that $\mathbf{k}^{*}$ should
be located at the boundary of the BZ. Furthermore, the point group of $%
\mathbf{k}^{*}$ should be only $C_{3}$, because $C_{6}$ will require that $%
\mathbf{k}^{*}$ is $\Gamma$ and $C_{4}$ is also ruled out for only $(\frac {1%
}{2},\frac{1}{2})$ in the BZ of square lattice owns $C_{4}$ symmetry and is
located at the boundary of the BZ (giving a non-vanishing $\mathbf{G}$ ),
however the displacement satisfying Eq. (\ref{delta}) will not let $C_{4}$
symmetry survive. For $C_{3}$, $\mathbf{k}^{*}$ can be $K$ or $K^{\prime}$
and because $K^{\prime}$ is related to $K$ by time-reversal symmetry,
hereafter we just consider $K$ shown in Fig.~S~\ref{sm}(c), which is equal
to $\frac{1}{3}\mathbf{b}_{1}+\frac{1}{3}\mathbf{b}_{2}$. Obviously, $%
c_{3}K=K-\mathbf{b}_{1}$, thus $\boldsymbol{\delta }=\mp\frac{1}{3}\mathbf{a}%
_{1}\pm\frac{1}{2}\mathbf{a}_{2}$. At last we require that the first layer
should be related to the third layer by the mirror plane coincided with the
second layer, which can guarantee that the coefficients before $q_{\mp}$ in $%
\mathcal{H}_{12}$ and $\mathcal{H}_{13}$ are the same.

As a matter of fact, we can focus on the lattice model shown in Fig.~S~\ref%
{sm}(a), whose primitive lattice basis vectors are denoted by $\mathbf{a}%
_{1} $ and $\mathbf{a}_{2}$, and in the Cartesian coordinate system shown in
Fig.~S~\ref{sm}(c), $\mathbf{a}_{1}=a(\frac{\sqrt{3}}{2},-\frac{1}{2})$ and $%
\mathbf{a}_{2}=a(0,1)$ where $a$ is the lattice parameter. Then we can get
the reciprocal lattice basis vectors, $\mathbf{b}_{1}=\frac{4\pi}{\sqrt 3a}%
(1,0)$ and $\mathbf{b}_{2}=\frac{4\pi}{\sqrt3a}(\frac{1}{2},\frac{\sqrt 3}{2}%
)$. The hexagonal BZ is shown in Fig.~S~\ref{sm}(c), and we have labeled
three points of symmetry, $\Gamma=(0,0)$, $M=\frac{1}{2}\mathbf{b}_{1}$ and $%
K=-K^{\prime}=\frac{1}{3}(\mathbf{b}_{1}+\mathbf{b}_{2})$. The three-layer
structure shown in Fig.~S~\ref{sm}(a) satisfies the requirements claimed
before: each layer is a triangular lattice and each unit cell only contains
one atom. According to such a prototypical structure, we then search the 3D
space groups from No. 143 to No. 194 and get the combinations of Wyckoff
positions which can give such a three-layer structure (or two) in a unit
cell, as listed in Table~S~\ref{wyckoff}. The combination like $(1a, 2h)$
means there is only one three-layer structure with the layer from $1a$ atoms
being the middle layer while the combination like $(2b, 4f)$ mean there are
two three-layer structures with $(2b)$ atoms forms two middle layers.

\begin{widetext}
\begin{table}
\centering
\begin{tabular}{|c|c|c|c|}
\hline
Space group & Combination of Wyckoff positions & Space group & Combination of Wyckoff positions\\
\hline
P312 & \begin{tabular}{c}
(1a, 2h), (1a, 2i), (1b, 2h), (1b, 2i) \\
(1c, 2g), (1c, 2i), (1d, 2g), (1d, 2i) \\
(1e, 2g), (1e, 2h), (1f, 2g), (1f, 2h)  \\
\end{tabular}
& P$6_3$22 & (2b, 4f), (2c, 4e), (2d, 4e), (2d, 4f) \\
\hline
P-31c & (2a, 4f), (2c, 4e), (2d, 4e), (2d, 4f) & P-6m2 & \begin{tabular}{c}
(1a, 2h), (1a, 2i), (1b, 2h), (1b, 2i) \\
(1c, 2g), (1c, 2i), (1d, 2g), (1d, 2i) \\
(1e, 2g), (1e, 2h), (1f, 2g), (1f, 2h)  \\
\end{tabular} \\
\hline
P-6 & \begin{tabular}{c}
(1a, 2h), (1a, 2i), (1b, 2h), (1b, 2i) \\
(1c, 2g), (1c, 2i), (1d, 2g), (1d, 2i) \\
(1e, 2g), (1e, 2h), (1f, 2g), (1f, 2h)  \\
\end{tabular} & P-6c2 & \begin{tabular}{c}
(2b, 4h), (2b, 4i), (2a, 4h), (2a, 4i) \\
(2c, 4g), (2c, 4i), (2d, 4g), (2d, 4i) \\
(2e, 4g), (2e, 4h), (2f, 4g), (2f, 4h)  \\
\end{tabular} \\
\hline
P$6_3$/m & (2a, 4f), (2c, 4e), (2d, 4e), (2d, 4f) & P-62c & (2b, 4f), (2c, 4e), (2d, 4e), (2d, 4f) \\
\hline
P$6_3$/mmc &  (2b, 4f), (2c, 4e), (2d, 4e), (2d, 4f) &  &  \\
\hline
\end{tabular}
\caption{The combination of different Wyckoff positions which will result in three iso-structure layers in one unit cell.}\label{wyckoff}
\end{table}\end{widetext}

\section{ Computational method}

The electronic band structure calculations have been carried out using the
full potential linearized augmented plane wave method as implemented in
WIEN2K package \cite{WIEN}. We used the Perdew-Burke-Ernzerhof (PBE)
generalized gradient approximation (GGA) exchange-correlation density
functional \cite{PBE}. A $12\times 12\times 1$\ mesh is used for the
Brillouin zone integral. A vacuum spacing of 20\AA\ is used so that the
interaction in the non-periodic directions can be neglected.